\RequirePackage{ifpdf}
\documentclass[12pt,letterpaper]{JHEP3}         
\usepackage{amssymb,amsfonts}

\usepackage{wrapfig}
\usepackage{epsfig}

\def\be{\begin{eqnarray}}
\def\ee{\end{eqnarray}}
\newcommand{\nn}{\nonumber}
\newcommand\para{\paragraph{}}

\newcommand{\eqn}[1]{(\ref{#1})}

\def\Dslash{\,\,{\raise.15ex\hbox{/}\mkern-12mu D}}
\def\Dbarslash{\,\,{\raise.15ex\hbox{/}\mkern-12mu {\bar D}}}
\def\delslash{\,\,{\raise.15ex\hbox{/}\mkern-9mu \partial}}
\def\delbarslash{\,\,{\raise.15ex\hbox{/}\mkern-9mu {\bar\partial}}}
\def\pslash{\,\,{\raise.15ex\hbox{/}\mkern-9mu p}}
\def\calDslash{\,\,{\raise.15ex\hbox{/}\mkern-12mu {\cal D}}}

\def\lae{\mathrel{\mathop{\smash{\lower .5 ex \hbox{$\stackrel<\sim$}}}}}
\def\lae{\mathrel{\mathop{\smash{\lower .5 ex \hbox{$\stackrel>\sim$}}}}}

\title{The Quantum Hall Effect in Supersymmetric Chern-Simons Theories}

\author{David Tong and Carl Turner\\
Department of Applied Mathematics and Theoretical Physics, \\
University of Cambridge, \\ 
Cambridge, CB3 OWA, UK \\
{\tt  d.tong, c.turner@damtp.cam.ac.uk}\\}

\abstract{
We introduce a supersymmetric Chern-Simons theory whose low energy 
physics is that of the fractional quantum Hall effect. The supersymmetry 
allows us to solve the theory analytically. We quantise the 
vortices and, by relating their dynamics to a matrix model, show that 
their ground state wavefunction is in the same universality class as the 
Laughlin state. We further construct coherent state representations of 
the excitations of a finite number of vortices. These are quasi-holes. 
By an explicit computation of the Berry phase, without resorting to a 
plasma analogy, we show that these excitations have fractional charge 
and spin. }

\begin{document}
\pagestyle{plain} \setcounter{page}{1}
\newcounter{bean}
\baselineskip16pt \setcounter{section}{0}

\section{Introduction and Summary}

Supersymmetry is a much beloved tool of high energy theorists. 
Supersymmetric field theories are often tractable, even at strong 
coupling, yet remain rich enough to exhibit a wide range of interesting 
dynamics.

\para 
In contrast, supersymmetric theories are much less studied in the condensed matter community, even in the limited role of toy 
models for strongly coupled phenomena. In part this is because 
supersymmetry typically provides analytic control for relativistic 
theories at vanishing chemical potential. At finite density, where most 
problems of interest in condensed matter lie, supersymmetry is usually 
broken and any advantage it brings is lost\footnote{There are a number of notable exceptions, including the role of supersymmetry in disorder \cite{efetov}, the possibility of emergent supersymmetry \cite{yuyang,sungsik,grover,bradlyn} and the study of supersymmetry protected phases \cite{schoutens}.}.

\para
There is, however, a class of theories in $d=2+1$ dimensions which are supersymmetric, yet non-relativistic \cite{llm}.  In these theories, supersymmetry is retained even at finite density. Despite the vast literature on supersymmetric field theories, the quantum dynamics of these models remains relatively unexplored. The purpose of this paper is to show that the low-energy physics of these theories is that of the fractional quantum Hall effect.

\para
Of course, the fractional quantum Hall effect is one of the most studied topics in physics over the past three decades. The theory rests on a beautiful and  intricate web of ideas involving microscopic wavefunctions \cite{laughlin}, low-energy effective Chern-Simons theories \cite{zhk,read,blokwen,fradkinl,bs} and boundary conformal theories \cite{chiralll,mr}. The supersymmetric theory that we present here is unlikely to be of direct relevance to any material. Instead, it   should be viewed as a toy model whose role is to highlight some of the links between these  different approaches to the quantum Hall effect. 

\para
In the rest of this introduction, we describe the supersymmetric model in more detail and explain what it's good for.  It is an Abelian Chern-Simons theory, coupled to both bosonic and fermionic non-relativistic matter fields. In this manner, it is an amalgamation of effective theories of \cite{blokwen} and \cite{fradkinl}. The model has vortices and these are viewed as the ``electrons". The vortices are ``BPS objects" \cite{wittenolive}: this means that they experience no classical static forces. It also means that they are protected by supersymmetry in a way which we describe in the main text.  This property allows us to perform an explicit quantisation of the vortex dynamics. We show that the  ground state wavefunction of the  vortices lies in the same universality class as the Laughlin wavefunction. It has the same long range correlations, but differs on short distance scales.

\para
We also describe the excitations of a droplet of vortices.  There are gapless, chiral edge excitations which, we show, are governed by the usual action for a chiral boson \cite{jf}, suitably truncated due to the presence of a finite number of vortices. Finally, we construct the quasi-hole excitations in this model and compute their Berry phase. This is, of course, a famous computation for the Laughlin wavefunctions \cite{frank}. However the usual analysis relies on the plasma analogy \cite{laughlin}, and the (admittedly well justified) assumption that the classical 2d plasma exhibits a screening phase. In contrast, here we are able to perform the relevant overlap integrals analytically to show that the quasi-holes have the expected fractional charge and statistics.

\para
Many of the properties of vortices described above follow from the fact that their dynamics is governed by a quantum mechanical matrix model, which was introduced by Polychronakos to describe quantum Hall physics \cite{alexios} and further studied in a number of works \cite{hellvram,ks,ks2,unknown}. The results of this paper show how this matrix model is related to more familiar effective field theories of the quantum Hall effect.

\para
The paper is organised as follows. In Section \ref{modelsec} we introduce the non-relativistic, supersymmetric theory. After a fairly detailed description of the symmetries of the theory, we discuss its two different phases and its spectrum of excitations.   Section \ref{vortsec} is devoted to a study of BPS vortices and contains the meat of the paper. We will show that the low-energy dynamics of vortices is governed by the matrix model introduced in \cite{alexios}. We review a number of results about this matrix model and derive some new ones. Finally, in  Section \ref{moresec} we offer some ideas for the future. A number of calculations are relegated to appendices.

\section{Non-Relativistic Chern-Simons-Matter Theories}\label{modelsec}

We start by introducing the $d=2+1$ non-relativistic, supersymmetric Chern-Simons theory. The theory consists of an Abelian gauge field $A_\mu$, coupled to complex scalar field $\phi$ and a complex fermion $\psi$.  The action is
\be S = \int dt d^2x && \left\{ i\phi^\dagger {\cal D}_0 \phi + i\psi^\dagger {\cal D}_0\psi - \frac{1}{2m}{\cal D}_\alpha \phi^\dagger\, {\cal D}_\alpha \phi - \frac{1}{2m}{\cal D}_\alpha \psi^\dagger\,{\cal D}_\alpha \psi  -\frac{k}{4\pi}  \epsilon^{\mu\nu\rho}A_\mu\partial_\nu A_\rho \right. \nn\\&&  \left. \ \ \ \ \ -\mu A_0  + \frac{1}{2m}\psi^\dagger B \psi 
- \frac{\pi}{mk}\left( |\phi|^4 - \mu |\phi|^2 + 3 |\phi|^2|\psi|^2\right)\right\}\label{lag1}\ee
%
Some conventions: the subscripts $\mu,\nu,\rho=0,1,2$ run over both space and time indices, while $\alpha=1,2$ runs over spatial indices only. The fermion carries no spinor index. Both $\phi$ and $\psi$ are assigned charge 1, so the covariant derivatives read ${\cal D}_\mu \phi =\partial_\mu \phi - i A_\mu \phi$ and similarly for $\psi$. The magnetic field is $B= \partial_1 A_2-\partial_2A_1$. Finally $|\psi|^2 = \psi^\dagger \psi =-\psi\psi^\dagger$.

\para
There are three parameters in the Lagrangian: the Chern-Simons level $k\in {\bf Z}^+$, the mass $m$ of both bosons and fermions, and the chemical potential $\mu$. As we will see later, the chemical potential $\mu$ can be more fruitfully thought of as a background magnetic field for vortices.  

\para
The first order kinetic terms mean that the action \eqn{lag1} describes bosonic and fermionic particles, but no anti-particles. The quartic potential terms correspond to delta function contact interactions between these particles. In the condensed matter context, the gauge field is considered to be emergent. One of its roles is to attach flux to particles through the Gauss' law constraint, which arises as the equation of motion for $A_0$,
\be B = \frac{2\pi}{k}\left(|\phi|^2 +|\psi|^2 - \mu\right)\label{gauss1}\ee
We'll learn more about the importance of this relation later.

\para
The action \eqn{lag1} can be constructed by starting from a  relativistic Chern-Simons theory with ${\cal N}=2$ supersymmetry and taking a limit in which the anti-particles decouple. For the case $\mu=0$, this was first done in \cite{llm} and we review the procedure in Appendix \ref{appa}.   To our knowledge, the supersymmetric  theory with $\mu\neq 0$ has not been previously constructed, although the bosonic sector of our theory is similar, but not identical, to a model studied by Manton \cite{manton} which shares the same vortices as \eqn{lag1}. We will describe these vortices in some detail in Section \ref{vortsec}. 

\subsection{Symmetries}

The action \eqn{lag1} is invariant under several symmetries. A number of these  play an important role in what follows and we provide the relevant details here.

\subsubsection*{Bosonic Symmetries}

Invariance under time translations gives rise to the Hamiltonian. After imposing the Gauss' law constraint \eqn{gauss1}, this takes the concise form
\be H = \frac{2}{m} \int d^2x \ \,|{\cal D}_z\phi|^2 + |{\cal D}_{\bar{z}}\psi|^2 +\frac{\pi}{k} |\phi|^2|\psi|^2 
\label{ham1}\ee
where $z=x^1+ix^2$ and $\bar{z}= x^1-ix^2$. Correspondingly,  $\partial_z = \frac{1}{2} (\partial_1 - i\partial_2)$ and $\partial_{\bar{z}}=\frac{1}{2}(\partial_1+i\partial_2)$. 

\para
Invariance under spatial translations gives rise to the complex momentum, $P=\frac{1}{2}(P_1-iP_2)$, which we write as
\be P = \hat{P} - \frac{\mu }{2} \int d^2x\ \bar{z}B\ \ \ \ {\rm with}\  \ \ \ \hat{P} = \int d^2x\ \phi^\dagger {\cal D}_z\phi - {\cal D}_z\psi^\dagger\psi
\label{mom}\ee
%
The $\hat{P}$ contribution is the standard Noether charge for spatial translations. The second term, proportional to the chemical  potential $\mu$, requires some explanation. As shown in  \cite{mantonnasir}, it arises because a translation is necessarily accompanied by a gauge transformation so that, for example,  $\delta_i\phi = {\cal D}_i\phi$. The presence of the chemical potential term $\mu A_0$ in the action then means that naive Noether charge for translations is not gauge invariant. This is remedied by the addition of a total derivative, resulting in the improved, gauge invariant momentum above. Note, however, that the resulting momentum $P$ is not itself translationally invariant. We shall comment further on this below.

\para
A similar subtlety occurs for rotations. The conserved angular momentum is given by
%
%
\be {\cal J} = \int d^2x\ \Big(z\phi^\dagger {\cal D}_z\phi - \bar{z} {\cal D}_{\bar{z}}\phi^\dagger\,\phi + z\psi^\dagger {\cal D}_z\psi - \bar{z} {\cal D}_{\bar{z}}\psi^\dagger\,\psi 
-\frac{1}{2}\psi^\dagger\psi -  \frac{\mu}{2}|z|^2B\Big) \label{J1}\ee
%
%
Here the first  terms are standard.   The penultimate term, $|\psi|^2$, arises because the fermion is taken to have spin $1/2$. The final term again arises as an improvement term in the Noether procedure which ensures that the resulting angular momentum is gauge invariant \cite{mantonnasir}. 


\para
The number of bosons and fermions in our model are individually conserved. The corresponding Noether charges are 
\be {\cal N}_B = \int d^2x\  \phi^\dagger\phi \ \ \ {\rm and}\ \ \  {\cal N}_F = \int d^2x\ \psi^\dagger \psi\label{nbnf}\ee
The total particle number is simply the charge under the Abelian gauge group
\be {\cal N} = {\cal N}_B + {\cal N}_F\nn\ee
We denote the axial combination as
\be {\cal R} = {\cal N}_B - {\cal N}_F\nn\ee
This will play the role of an R-symmetry in the supersymmetry algebra.

\para
The presence of the anomalous term in the expression for the momentum \eqn{mom} has an interesting effect on the commutation relations. 
(Here we describe the quantum commutation relations rather than classical Poisson brackets.) We find
\be [H,\hat{P}] = -\frac{2\pi \mu}{mk}\, \hat{P}\ \ \ \ {\rm and}\ \ \ \ [H,P]=0\label{updown}\ee
So the Noether charge $P$ is conserved, but the translationally invariant momenta $\hat{P}^\dagger$ and $\hat{P}$ act as raising and lowering operators for the spectrum. Further, the conserved momenta do not commute. We have
\be [P,P^\dagger ] = - \frac{\pi  \mu}{k} {\cal N}\label{roundandround}\ee
Both \eqn{updown} and \eqn{roundandround} are similar to  the commutation relations in quantum mechanics for momenta in a magnetic field. This is because, as we will describe in more detail below, $\mu$ acts like an effective magnetic for vortices while  the Gauss' law constraint ensures that all excitations carry some vortex charge. 

\para
When $\mu=0$  the theory also enjoys both a Galilean boost  and, more surprisingly, a (super)conformal symmetry  \cite{llm} (see also \cite{horvathy1}). In this paper, we restrict ourselves to the non-conformal theory with chemical potential $\mu >0$.

\subsubsection*{Supersymmetries}

The action \eqn{lag1} enjoys two complex supersymmetries \cite{llm}.  These often go by the name of {\it kinematical} and {\it dynamical} supersymmetries.
The kinematical supersymmetry is the simpler of the two. 
\be \delta_1 \phi = \epsilon_1^\dagger \psi\ \ \ ,\ \ \ \delta_1\psi = -\epsilon_1 \phi\ \ \ ,\ \ \  \delta_1 A_z =0 \ \ \ ,\ \ \ \delta_1A_0 = \frac{\pi}{mk}\left(\epsilon_1\phi\psi^\dagger - \epsilon_1^\dagger \psi\phi^\dagger\right)\ \ \ \ \ \nn\ee
%
This is reminiscent of the Green-Schwarz spacetime supersymmetry on the string worldsheet.
The transformation on $\phi$ and $\psi$ is generated by
\be Q_1 = \int d^2x \ \phi^\dagger \psi\label{Q1}\ee
This does not specify a transformation for $A_0$ which does no harm as long as we allow ourselves to impose Gauss' law. We will see the implications of this below. 

\para
Under the dynamical supersymmetry, the fields transform as
\be \delta_2\phi = \epsilon_2^\dagger {\cal D}_{\bar{z}}\psi&&\ \ \! ,\ \ \ \delta_2\psi = \epsilon_2{\cal D}_z\phi\ \ \ ,\ \ \  \delta_2 A_z = -\frac{i\pi}{k}\epsilon_2^\dagger \psi\phi^\dagger\nn\\ \delta_2 A_0 &=& \frac{i \pi}{mk }\left(\epsilon_2^\dagger\phi^\dagger {\cal D}_{\bar{z}}\psi -\epsilon_2\phi{\cal D}_z\psi^\dagger \right)\nn\ee
%
with supercharge 
\be Q_2 =  \int d^2x\ \phi^\dagger {\cal D}_{\bar{z}}\psi\label{Q2}\ee
The supersymmetry algebra is 
\be \{Q_1,Q_1^\dagger\} = {\cal N}\ \ \ ,\ \ \ \{Q_2,Q_2^\dagger\} = \frac{m}{2}H\ \ \ , \ \ \ \{Q_1,Q_2^\dagger\} =  \hat{P}\label{susyalg}\ee
Note that the two supercharges generate the translationally invariant momentum $\hat{P}$, rather than the conserved momentum $P$. There is also a mild surprise in the commutators of bosonic and fermionic charges\footnote{We thank Nima Doroud for very useful discussions regarding this algebra.}, in particular
\be [H,Q_1] =  -\frac{2\pi\mu}{mk} \,Q_1\label{oddcomm}\ee
This means that although the kinematic supersymmetries leave the action invariant, when $\mu\neq 0$ they do not result in a symmetry of the spectrum. This can be traced to the fact that Gauss' law was required, both in the construction of the Hamiltonian \eqn{ham1} and in the derivation of the commutators \eqn{oddcomm}. (To our knowledge, the possibility of non-relativistic supersymmetry generator $Q_1$ was first raised in \cite{clarklove} who also pointed out that this generator is spontaneously broken in any vacuum with non-vanishing particle number.) Other commutators follow from Jacobi identities and give $[Q_2,H] = [Q_1,\hat{P}] = [Q_1^\dagger,\hat{P}] = 0$ while $[Q_2,\hat{P}] = (2/m) [H,Q_1]$.

\para
Finally, the  commutators of the angular momentum will also be important for our story. There's nothing unusual about them. We have
\be [{\cal J},Q_1] = -\frac{1}{2} Q_1\ \ \ {\rm and}\ \ \ \ [{\cal J},Q_2] = \frac{1}{2}Q_2\label{jq}\ee
 which is the statement that the supercharges have spin $\mp 1/2$. The total particle number commutes with all supersymmetries, $[{\cal N},Q_\alpha]=0$, but only because of cancellations between boson and fermion numbers. Individually, we have
\be [{\cal N}_B, Q_\alpha] =  -Q_\alpha\ \ \ {\rm and}\ \ \ [{\cal N}_F,Q_\alpha ] = +Q_\alpha\nn\ee
%
This justifies our previous claim that ${\cal R}$ is an ``R-symmetry" since $[{\cal R},Q_\alpha] = 2Q_\alpha$. From these, we deduce that 
\be [{\cal J}-\frac{1}{2}{\cal N}_F,Q_2]=0\label{jisinvariant}\ee
This fact will be important in Section \ref{trapsec}.


\subsection{The Vacuum, The Hall Phase, and Excitations}\label{phasessec}

We now describe some basic features of the dynamics of our model. Because non-relativistic field theories have no anti-particles, the theory decomposes into sectors labelled by the conserved particle numbers which, in our case, are ${\cal N}_B$ and ${\cal N}_F$. To solve the theory, we need to determine the energy spectrum in each of these sectors.  

\para
One way to organise these sectors is to start with the ${\cal N}=0$ Hilbert space and build up by adding successive particles. Instead, we will take a dual perspective. Our theory enjoys a topological current, 
\be J^\mu = \frac{1}{2\pi} \epsilon^{\mu\nu\rho}\partial_\nu A_\rho\label{vcurrent}\ee
The associated particles are vortices. We will view these vortices as the ``electrons" of our theory. 

\para
Our theory has two translationally invariant ground states  consistent with Gauss' law \eqn{gauss1}, both of which have $H=0$. We call these the {\it vacuum} and the {\it Hall Phase}. They are given by
\be \mbox{\underline{The Vacuum:}}&&\ \ \ \ \ \ \  |\phi|^2 =\mu\ \ \ {\rm and}\ \ \ B=0\label{classicalvac}\\ \ \nn\\
 \mbox{\underline{The Hall Phase}:}&&\ \ \ \ \ \ \  |\phi|^2 =0 \ \ \ {\rm and}\ \ \ B=-\frac{2\pi\mu}{k}\label{hallphase}\ee
The vacuum state contains no vortices, $\int d^2x\ J^0=0$. However, the bosons have condensed which means that the particle number is ${\cal N}=\infty$. In contrast, the Hall phase has vanishing particle number but infinite vortex number, $\int d^2x\ J^0=\infty$. 

\para
The purpose of this paper is to understand what happens as we inject vortices into the vacuum. For any finite number of vortices, the system  breaks translational invariance. But, as we fill the plane with vortices, the Hall phase emerges. In Section \ref{vortsec}, we tell  both the classical and quantum versions of this story in some detail. First, however, we describe some simple properties of excitations above each of these ground states.

\subsubsection*{The Vacuum}

%
%

%
%

The  key feature of the vacuum state is that $U(1)$ gauge symmetry is broken. This  ensures that the theory admits topological, localised vortex solutions. These vortices will be the main focus of this paper and we postpone a more detailed discussion until Section \ref{vortsec}. 
Here, we summarise their three main properties:
\begin{itemize}
\item Vortices are gapless. States with an arbitrary number of vortices exist with $H=0$.
\item Vortices have statistical phase $\pi k$. This means that the vortices are bosons when $k$ is even and fermions when $k$ is odd.
\item Vortices are singlets under supersymmetry.
\end{itemize}

There are further excitations above the vacuum arising from the fundamental fields $\phi$ and $\psi$. These excitations are both gapped, with an excitation energy $2\pi \mu /mk$. These excitations can be generated from the vacuum by using the raising operators $\hat{P}^\dagger$ and $Q_1^\dagger$, together with the supercharges $Q_2$ and $Q_2^\dagger$.

\subsubsection*{The Hall Phase}

%
%

The Hall phase  has an unbroken $U(1)$ gauge symmetry and the long-distance physics is dominated by the Chern-Simons term.  It is well known that such theories capture the essential properties  of the fractional quantum Hall effect. We now take the opportunity to review this standard material (see, for example, \cite{zee,wenreview} for reviews).  

\para
To describe quantum Hall physics, it is not enough to specify the Lagrangian; we need to know  how electromagnetism couples to the theory\footnote{There are two, dual, descriptions of the long-wavelength quantum Hall physics in terms of Chern-Simons theories. In one description, the Chern-Simons level is equal to $\nu$, the filling fraction \cite{zhk,read}, the electrons are the fundamental excitations and the vortices the fractionally charged quasi-particles. Here we are interested in the dual description, related by a particle-vortex duality transformation, where the Chern-Simons coefficient is $1/\nu$ and the electrons are vortices.}. (Recall that the Abelian gauge field $A_\mu$ in the Lagrangian \eqn{lag1} should be thought of as an emergent, statistical gauge field, not the electromagnetic field.) Since we wish to treat the vortices as the ``electrons" of the theory, the background electromagnetic field $A_\mu^{\rm ext}$ must couple to the topological current \eqn{vcurrent},
\be {\cal L}_{\rm Hall} = \frac{k}{4\pi}\epsilon^{\mu\nu\rho}A_\mu \partial_\nu A_\rho + eA^{\rm ext}_\mu J^\mu+ \ldots\nn\ee
Here $e$ denotes the electron charge, while $\ldots$ includes the rest of the Lagrangian \eqn{lag1}, as well as the $d=3+1$ dimensional Maxwell term for $A_\mu^{\rm ext}$. 

\para
We momentarily ignore the fundamental fields $\phi$ and $\psi$. Integrating out $A_\mu$, the quadratic Lagrangian for the background field is given by
\be {\cal L}_{\rm Hall} = -\frac{e^2}{4\pi k} \epsilon^{\mu\nu\rho} A_\mu^{\rm ext}\partial_\nu A_\rho^{\rm ext} + \ldots\nn\ee
The effective action $S_{\rm eff}[A^{\rm ext}] = \int d^3x\, {\cal L}_{\rm Hall}$, is now a functional of the non-dynamical, background  electromagnetic field. Its role is to tell us how the system responds to an applied electromagnetic field through the relation $\langle J^\mu \rangle = \partial {\cal S}_{\rm eff}/\partial A_\mu^{\rm ext}$. The result is a Hall conductivity, 
\be \sigma_H = \frac{e^2}{2\pi k}\label{hallcond}\ee
This is the response of a fractional quantum Hall fluid at filling fraction $\nu = 1/k$.

\para
\EPSFIGURE{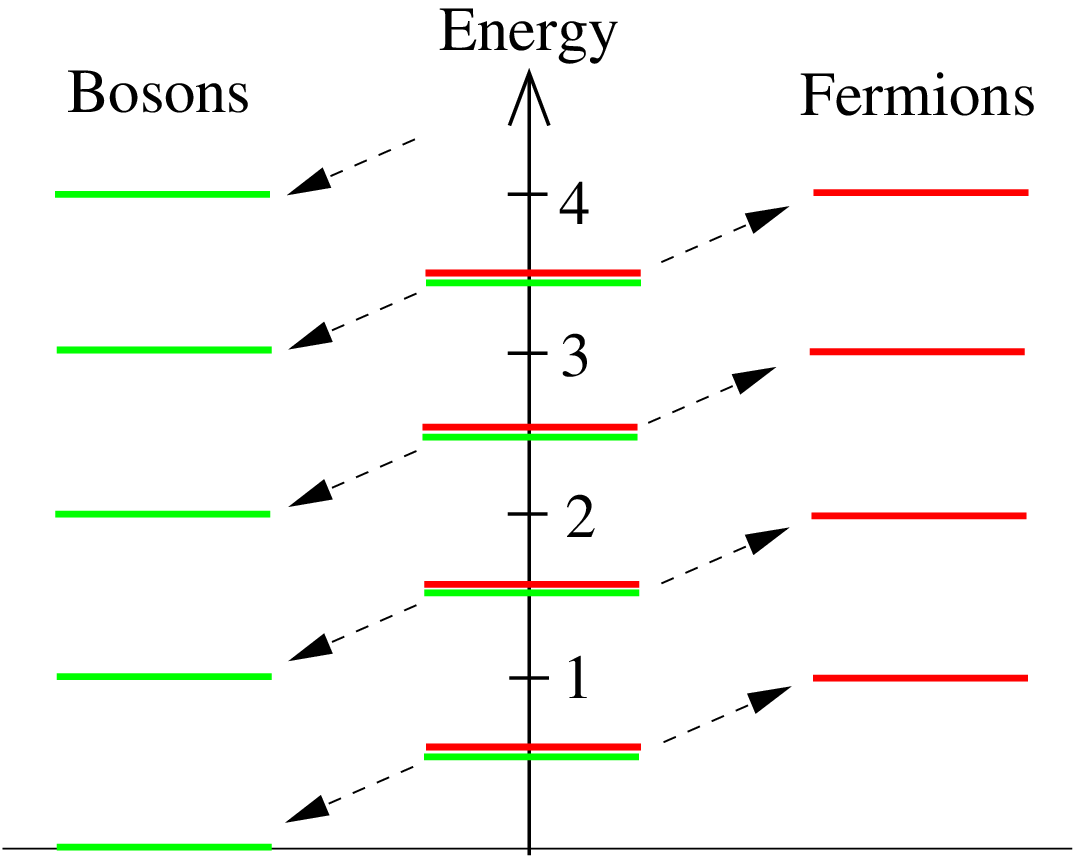,height=140pt}{Fundamental excitations}

Let us now return to the fundamental fields $\phi$ and $\psi$. Each of these experiences a magnetic field $B = - 2\pi\mu/k$ and forms Landau levels. The usual Landau level quantisation results in a spectrum 
\be 
E_{LL} = \frac{|B|}{m}(l + 1/2)\nn\ee
 with $l=0,1,\ldots$. However, the Lagrangian \eqn{lag1} also includes extra terms which shift the overall energy of these states. The shift is  down for bosons and up for fermions, as shown in the figure. The net result is that the energies of the Landau levels, at leading order, are given by
\be E = \frac{2\pi\mu l}{mk} \ \ \ \ \left\{\begin{array}{ll} l=0,1,2,\ldots\ \ \ \ &{\rm for}\ \phi\\ l=1,2,\ldots& {\rm for}\ \psi\end{array}\right.\nn\ee
The gapped states, with $l\geq 1$, arising from $\phi$ have spin $1/2k$; those arising from $\psi$ have spin $(1+k)/2k$. The Gauss law constraint \eqn{gauss1} ensures that, when coupled to a background electromagnetic field, each of these carries charge $-e/k$. These are the quasi-particle excitations of our supersymmetric quantum Hall fluid. The supercharges $Q_2$ and $Q_2^\dagger$ map between the fermionic and bosonic gapped Landau levels.

\para
The system also has an a gapless band of quasi-particles, arising from the lowest Landau level of $\phi$. These modes are not free: they interact through the $\phi^4$ potential in \eqn{lag1}. Nonetheless, supersymmetry ensures that these states have vanishing energy at all orders of perturbation theory. This is because the commutation relations for $Q_2$ require that any excitation with $H>0$ must be paired with an excitation that differs by spin $1/2$. Yet the states in  lowest Landau level have no partners and must, therefore,  remain at zero energy. In essence, the theory has an infinite Witten index ${\rm Tr}(-1)^F$.  
If we start from the lowest Landau level, we can build up to higher levels by acting with $\hat{P}^\dagger$ and $Q_1^\dagger$.

\para
The presence of a gapless Landau level may appear to contradict our claim that this system describes quantum Hall physics. After all, one of the defining features of a quantum Hall state is that it is gapped and incompressible. We will resolve this in Section \ref{vortsec} by studying how the Hall phase emerges from vortices when placed in a confining potential. 
We will show that, for any finite number of vortices, there is a unique incompressible droplet of lowest angular momentum. However, in the absence of a confining potential, this droplet has zero energy edge modes and zero energy quasi-hole excitations. The gapless Landau level describes these degrees of freedom for an infinite number of BPS vortices, an interpretation recently suggested in  a different context in  \cite{ken}.  We will revisit this in Section \ref{edgesec} in the context of the non-commutative approach to quantum Hall physics.

\para
It is worth mentioning that this situation is not unusual in quantum Hall systems. The special, ultra-local Hamiltonians (such as Haldane pseudo-potentials) used as models of quantum Hall physics also have zero energy edge modes and zero-energy quasi-hole excitations for finite droplets. See, for example, \cite{mil,readchiral} for related discussions.


\section{A Quantum Hall Fluid of Vortices}\label{vortsec}

We would like to understand  how to interpolate from the vacuum to the Hall phase. We do this by injecting vortices.  These vortices are BPS which, in this context, means that they have $H=0$ and lie in a  protected sector of the theory.  From the form of the Hamiltonian \eqn{ham1} and Gauss' law \eqn{gauss1}, it is clear that solutions with vanishing energy, $H=0$, can be constructed by solving the equations 
\be {\cal D}_z\phi =0\ \ \ {\rm and}\ \ \ B= \frac{2\pi}{k}(|\phi|^2 - \mu)\label{vortex}\ee
with the fermions set to zero: $\psi=0$.

\para
The vortex equations \eqn{vortex} are well studied. Solutions are labelled by the integer winding of the scalar field $\phi$ or, equivalently, by the magnetic flux
\be n = -\frac{1}{2\pi}\int d^2x\ B \ \in\  {\bf Z}^+\label{fluxnumber}\ee
In the sector with winding $n$, the most general solution to \eqn{vortex} has $2n$ real parameters \cite{erick,taubes}. These parameters are referred to as {\it collective coordinates} or, in the string theory literature, {\it moduli}. 
When vortices are well separated, these correspond to $n$  positions on the complex plane.  The existence of these moduli reflects the fact that the coefficient of the quartic interaction in \eqn{lag1}  has been tuned to the critical value, ensuring that there are neither attractive nor repulsive forces between the vortices. 

\begin{figure}[!h]
\begin{center}
\includegraphics[height=0.7in]{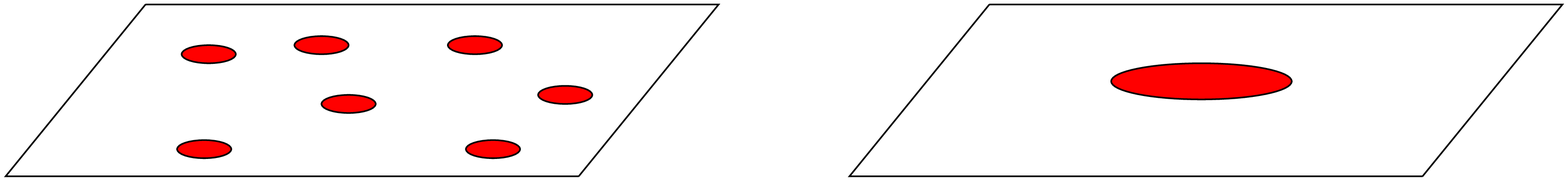}
\end{center}
\caption{Two points in the moduli space of $n=7$ vortices}\label{vortexfig}
\end{figure}
\noindent

\para
As vortices coalesce, they lose their individual identities and the interpretation of these moduli changes. It is tempting to label the vortex by the point at which the Higgs field vanishes, but this does not provide an accurate description of what the vortex profile looks like. Instead, as we show in Section \ref{edgesec}, in this regime it is better to think of the $2n$ moduli as describing the edge modes of  a large, incompressible fluid.


\subsubsection*{Why do Vortices Form a Fractional Quantum Hall State?}

The rest of this section is devoted to a detailed analysis of the quantum dynamics of vortices. We will ultimately show that their ground state is given by the Laughlin wavenfunction. But here we first provide a hand-waving argument for why we expect the vortices to form a quantum Hall fluid.

\para
We first note that the chemical potential term $\mu A_0$, present in the Lagrangian \eqn{lag1}, can be viewed as a background magnetic field for vortices. It can be written as 
\be -\int d^3 x\ \mu A_0=\int d^3x\ eJ^\mu A_\mu^{\rm ext} \nn\ee
where $J^\mu$ is the topological current \eqn{vcurrent} and 
\be 
A_\alpha^{\rm ext} = -\frac{B^{\rm ext}}{2}\epsilon_{\alpha\beta}x^\beta  \ \ \ \ {\rm with}\ \ \ \ B^{\rm ext} = \frac{2\pi\mu}{e}
\label{bext}\ee
This means that we expect the dynamics of vortices to correspond to particles moving in a background magnetic field. Nonetheless, it may be rather surprising that the vortices form a Hall state because, as we have seen, there is no force between the vortices. Yet the key physics underlying the fractional quantum Hall effect is the repulsive interactions between electrons, opening up a gap in the partially filled Landau level.

\para
Although there is no force between vortices, they are not point particles. Instead, they are solitons obeying non-linear equations and, as they approach, the solutions deform. Indeed, when the vortices are as closely packed as they can be, they form a classically incompressible fluid  as shown in the right-hand side of Figure \ref{vortexfig}.  The scalar field $\phi$ has an $n^{\rm th}$ order zero in the centre of the disc and numerical  studies show that the solution is well approximated as a disc of magnetic flux in which $\phi=0$ and  $B=-2\pi \mu/k$. This has motivated the ``bag model" of vortices in \cite{stefano1,stefano2}. 
For us, it means that the vortex is a  droplet of what we have called the ``Hall phase". 

\para
When $n$ vortices coalesce, the radius $R$ of the resulting droplet can be estimated using the flux quantisation \eqn{fluxnumber} to be 
\be R \approx \sqrt{\frac{kn}{\pi\mu}}\label{R}\ee
Now we can do a back-of-the-envelope calculation. In a magnetic field $B^{\rm ext}$, the number of states per unit area in the lowest Landau level is $eB^{\rm ext}/2\pi=\mu$. In an area $A=\pi R^2 = nk/\mu$, the lowest Landau level  therefore admits $B^{\rm ext} A = nk$ states. We've placed $n$ vortices in this region, so the filling fraction is 
\be \nu = \frac{1}{k}\nn\ee
This, of course, is the expected filling fraction in the Hall phase with conductivity \eqn{hallcond}.
\be \frac{eB^{\rm ext}}{2\pi} = \mu\ \ \ \ \ \ \ \pi R^2 = \frac{nk}{\mu}\nn\ee

\subsection{The Dynamics of Vortices}

We now turn to a more detailed description of the dynamics of vortices. We first introduce the {\it vortex moduli space}, ${\cal M}_n$. This is space of solutions to the vortex equations \eqn{vortex} with winding number $n$. As we have already mentioned, 
\be {\rm dim}({\cal M}_n) = 2n\nn\ee
The coordinates $X^a$, $a=1,\ldots, 2n$, parameterising ${\cal M}_n$ are the collective coordinates of vortex solutions: $\phi(x;X)$ and $A_\alpha(x;X)$. 
\para
The standard approach to soliton dynamics is to assume that, at low-energies, motion can be modelled by restricting to the moduli space \cite{manmon}. This is usually applied in relativistic theories where the action is second order in time derivatives and typically provides an accurate approximation to the real dynamics. Here we have  a non-relativistic theory, first order in time derivatives, and this results in a number of differences which we now explain.
One ultimate surprise --- which we will get to in Section \ref{trapsec} --- is that there is no approximation involved in the moduli space dynamics in this system; instead it is exact. 

\para
The first, and most important difference, is associated to the meaning of the space ${\cal M}_n$. In relativistic theories, ${\cal M}_n$  is the configuration space of vortices and the dynamics is captured by geodesic motion on ${\cal M}_n$ with respect to a metric $g_{ab}(X)$. It is known that ${\cal M}_n$ is a complex manifold, with complex structure $J$, and the metric $g_{ab}(X)$ is K\"ahler. For completeness, we explain how to construct this metric in Appendix \ref{appgeom}.

\para
In our non-relativistic context, it is no longer true that ${\cal M}_n$ is the configuration space of vortices. Instead, it is the {\it phase space}. The dynamics of the vortices is described by a quantum mechanics action of the form,
\be S_{\rm vortex} = \int dt\ {\cal F}_a(X)\dot{X}^a\label{1vdynamics}\ee
where ${\cal F}(X)$ is a one-form over ${\cal M}_n$. Our goal is determine this one-form.

\para
In fact, this problem has already been solved in the literature. A model which shares its vortex dynamics with ours was previously studied by Manton \cite{manton} and subsequently, in more geometric form, in \cite{romao,nunomartin}. The main result of these papers is that ${\cal F}$ is an object known as the symplectic potential. It has the property that
\be d{\cal F} = \Omega\label{kform}\ee
where $\Omega$ is the K\"ahler  form on ${\cal M}_n$, compatible with the metric $g_{ab}$ and the complex structure $J$. 

\para
For a single vortex, the moduli space is simply the plane ${\bf C}$ and the K\"ahler form is 
\be \Omega = \frac{\pi \mu}{2}\,dz\wedge d\bar{z}\nn\ee
For  $n\geq 2$ vortices the K\"ahler form is more complicated. 
We describe the construction of $\Omega$ in Appendix \ref{appgeom}. Explicit expressions are only known for well-separated vortices \cite{nunomartin}. 

\para
The derivation of \eqn{kform} given in \cite{manton,romao,nunomartin} relies on a parameterisation of the vortex moduli space introduced  earlier in \cite{samols}. The use of these coordinates means that the calculation is not entirely straightforward. For this reason, in Appendix \ref{appgeom}, we present a simpler derivation of \eqn{kform} which does not rely on any choice of coordinates. (For a different approach to particle dynamics appropriate for vortices, see \cite{horvathy2}.)

\subsubsection*{There are no Fermion Zero Modes}

The vortices are BPS states: they are annihilated by the supercharge $Q_2$. In the context of first order dynamics, this means that the collective coordinates $X$ do not transform under $Q_2$. In particular, there are no accompanying Grassmann collective coordinates. Indeed, it is simple to check explicitly that there are no fermionic zero modes in the background of the vortex. 

\para
The upshot is that the vortices themselves are supersymmetric singlets. The role of supersymmetry in the vortex dynamics \eqn{1vdynamics} is restricted to ensuring the vortices have strictly vanishing energy, $H=0$, in the full quantum theory.

\para
The fact that the BPS solitons have no fermion zero modes may come as something of a surprise. Indeed, it is rather different from what happens for BPS solitons in relativistic field theories or in string theory. It is worth pausing to explain this difference. In more familiar relativistic theories, if a soliton is invariant under a given supercharge ${\cal Q}$ then that supercharge will descend to the worldvolume theory, relating bosonic and fermionic zero modes on the worldvolume. However, when we say that a soliton is invariant under ${\cal Q}$, we mean that the static configuration is invariant: when the soliton moves, the supercharge ${\cal Q}$ typically acts and generates a fermionic zero mode. This means that while ${\cal Q}$ does not act on the bosonic configuration space of the soliton, it does act on the phase space. 

\para
In our non-relativistic theory, the statement that $Q_2$ annihilates the soliton is stronger: it  means that $Q_2$ does not act on the soliton phase space. This is the reason that there are no associated fermionic zero modes. 

\subsection{Introducing a Harmonic Trap}\label{trapsec}

We have derived a low-energy effective action \eqn{1vdynamics} for the vortex dynamics. However, this dynamics is boring. The equation of motion arising from \eqn{1vdynamics} is 
\be \Omega_{ab}\dot{X}^b = 0 \ \ \ \Rightarrow \ \ \ \dot{X}^a = 0 \nn\ee
The vortices don't move. They are pinned in place. 

\para
The lack of dynamics follows because there is no force between vortices and, in a first order system, we don't have the luxury of giving the vortices an initial velocity. To get something more interesting, we impose an external force on the vortices. We will do so by introducing a harmonic trap. We want this trap to be compatible with supersymmetry. We can do this by choosing the new Hamiltonian
\be H_{\rm new} = H +\omega\left({\cal J} -\frac{1}{2}{\cal N}_F\right)\nn\ee
where ${\cal J}$ is the angular momentum \eqn{J1}, ${\cal N}_F$ the fermion number operator \eqn{nbnf} and $\omega$ dictates the strength of the trap.  From \eqn{jisinvariant}, we see that this Hamiltonian remains invariant under $Q_2$, although not $Q_1$. When evaluated on BPS vortices, the Hamiltonian is simply
\be H_{\rm new} = -\frac{\mu\omega}{2}\int d^2x\ |z|^2 B\label{simpleham}\ee
This new Hamiltonian is the angular momentum of a given BPS vortex configuration: it preserves the BPS nature of vortices while shifting their energy.  Evaluating \eqn{simpleham} on a vortex configuration provides a  function ${\cal J}(X)$ over the vortex moduli space ${\cal M}_n$ which governs the their low-energy dynamics, 
\be S_{\rm vortex} = \int dt\ \left({\cal F}_a(X)\dot{X}^a -  \omega{\cal J}(X)\right)\label{vpotential}\ee
We will now look at some examples of the classical dynamics	 dcescribed by this action.

\subsubsection*{Classical Motion in the Trap}

The harmonic trap \eqn{simpleham} favours those vortex solutions that are clustered towards the origin.  The lowest energy configuration now has all vortices coincident at the origin, as in the right-hand picture in Figure \ref{vortexfig}.  As we have seen, the  size of this vortex is given by \eqn{R}, so the angular momentum of this state is 
\be {\cal J}_0  \approx - \frac{\mu}{2}  \int_0^R  dr\ 2\pi r^3 B = \frac{ k n^2}{2} \label{jmin}\ee
This is the only static configuration. 
All other solutions evolve through the equation of motion
\be \Omega_{ab}\dot{X}^b =\omega \frac{\partial {\cal J}}{\partial X^a}\label{whatxdoes}\ee
In particular, a single vortex displaced a distance $r\gg \sqrt{1/\mu}$ from the origin,  will have angular momentum ${\cal J} \approx \pi \mu r^2$. This vortex orbits around the origin with frequency $\omega$. 

\para
There is something rather surprising about the moduli space approximation for this first order dynamics: it is exact! The solutions to the equation of motion in the presence of the trap are simply time dependent rotations of the static solutions so, for example, $\phi = \phi(x;X(t))$, with $X(t)$ obeying \eqn{whatxdoes}. This a property of any first order system  with a Hamiltonian, such as $H={\cal J}$, which acts as a symmetry generator on the moduli space.

\subsection{The Quantum Hall Matrix Model}

The description of the vortex dynamics \eqn{vpotential} is, unfortunately, rather abstract. For $n\geq 2$ vortices, we have only implicit definitions of the K\"ahler form $\Omega$ and the angular momentum ${\cal J}$ on the vortex moduli space. It seems plausible that one could make progress using the parameterisation of the vortex moduli space introduced in \cite{samols}. Here, however, we take a different approach.

\para
An alternative description of the vortex moduli space is provided by D-branes in string theory \cite{me}. This is analogous to the ADHM construction of the instanton moduli space. The vortex moduli space ${\cal M}_n$ is parameterised by:
\begin{itemize}
\item An $n\times n$ complex matrix $Z$
 \item A $n$-component complex vector $\varphi$
\end{itemize}
These provide $n(n+1)$ complex degrees of freedom. We will identify configurations related by the $U(n)$ action
\be Z \rightarrow U Z U^\dagger\ \ \ {\rm and}\ \ \ \varphi \rightarrow U\varphi\ \ \ \ \ {\rm with}\ U \in U(n)\label{unaction}\ee
We further require that $Z$ and $\varphi$  satisfy the matrix constraint\footnote{As an aside: for  relativistic vortices, the right-hand side of \eqn{constraint} is $2\pi/e^2$, where $e^2$ is the gauge coupling constant. Comparing the vortex equations \eqn{vortex} to their relativistic counterparts shows that this becomes $k$ in the non-relativistic context. The fact that this is integer valued for vortices in the Chern-Simons theory will prove important below.},
\be \pi\mu\,[Z,Z^\dagger] + \varphi\varphi^\dagger = k\,{\bf 1}_n\label{constraint}\ee
This constraint is the moment map for the action \eqn{unaction} with level $k$. We define the moduli space $\tilde{\cal M}_n$ through the symplectic quotient,
\be  \tilde{M}_n = \left\{ Z, \varphi \ \ {\rm such\ that}\ \ \pi\mu [Z,Z^\dagger] +\varphi\varphi = k\right\} / U(n)\nn\ee
This space has real dimension ${\rm dim}(\tilde{\cal M}_n)=2n$. The string theory construction of \cite{me} shows that this space is related to the vortex moduli space
\be \tilde{\cal M}_n \cong {\cal M}_n\nn\ee
These spaces are conjectured to be isomorphic as complex manifolds, and have the same K\"ahler class. To our knowledge, there is no direct proof of this conjecture beyond the string theory construction provided in \cite{me}.  

\para
The matrix description provides a different parametrisation of the vortex moduli space. When the vortices are well separated, $Z$ is approximately diagonal. The positions of the vortices are described by these $n$ diagonal elements.  (The normalisation of $\pi\mu$ in \eqn{constraint} is associated to the vortex size.)  However, as the vortices approach, $Z$ is no longer approximately diagonal, reflecting the fact it is better to think of the  locations of the vortices as fuzzy, spread out over a disc of radius \eqn{R}. This feature is captured by the matrix description of the vortex moduli space. 

\para
The moduli space $\tilde{M}_n$ inherits a natural metric through the quotient construction described above. This does {\it not} coincide with the metric on the vortex moduli space ${\cal M}_n$ described in Appendix \ref{appgeom}. Nonetheless, there are now a number of examples in which computations of BPS quantities using $\tilde{\cal M}_n$ coincide with those of computed from the vortex moduli space ${\cal M}_n$ because they are insensitive to the details of the metric (see, for example, \cite{meami,tudor,yoshida,field}). Here we will ultimately be interested in holomorphic wavefunctions over the vortex moduli space. Assuming the conjectured equivalence of the spaces as complex manifolds, it will suffice to work with the matrix model description of the vortex moduli space.

\subsubsection*{The Matrix Model Action}

It is now a simple matter to write the vortex dynamics in terms of these new fields. We introduce a $U(n)$ gauge field, $a_0$, on the worldline of the vortices. In the absence of a harmonic trap, the low-energy vortex dynamics is governed by the $U(n)$ gauged quantum mechanics,
\be S_{\rm vortex} = \int dt\ i\pi\mu\,{\rm Tr}\,\left(Z^\dagger {\cal D}_0 Z\right) + i \varphi^\dagger {\cal D}_0\varphi - k\,{\rm Tr}\,a_0
\label{polyact}\ee
where ${\cal D}_0 Z = \partial_0 Z - i[a_0,Z]$ and ${\cal D}_0 \varphi = \partial_0\varphi - ia_0\varphi$. The quantum mechanical Chern-Simons term ensures that Gauss' law for the matrix model coincides with \eqn{constraint}. This means that this action describes the same physics as \eqn{1vdynamics}.

\para
The action \eqn{polyact} is the  {\it quantum Hall matrix model}, previously proposed as a description of the fractional quantum Hall effect by Polychronakos \cite{alexios} and further explored in \cite{hellvram,ks,hkk,cappelli,cappelli2,edge}. The connection to first order vortex dynamics was  noted earlier in \cite{unknown}.

\para
We note in passing that we've used the D-brane construction of \cite{me} in a fairly indirect way to derive the quantum Hall matrix model. A more direct D-brane derivation of the matrix model was provided previously  in \cite{oren}. It would be interesting to see how this work, or the string theory construction of \cite{tadashi}, is related to the present set-up.

\para
We would also like to add the harmonic trap to the matrix model. This too was explained in \cite{alexios}. Spatial rotation within the matrix model acts as $Z\rightarrow e^{i\theta}Z$, with the associated charge ${\cal J} = \pi\mu{\rm Tr} \,Z^\dagger Z$. Adding this to the action, we get the matrix model generalisation of \eqn{vpotential},
\be S_{\rm vortex} = \int dt\ i\pi\mu\,{\rm Tr}\,\left(Z^\dagger {\cal D}_0 Z\right) + i \varphi^\dagger {\cal D}_0\varphi - k\,{\rm Tr}\,a_0 - \omega \pi\mu\,{\rm Tr}\left(Z^\dagger Z\right)\label{qhmm}\ee
In the rest of this section, we describe  the properties of this matrix model. Much of this is review of earlier work, in particular \cite{alexios} and \cite{hellvram,ks}. However, we also make a number of new observations about the matrix model, most notably the computation of the charge and statistics of quasi-hole excitations.

\subsubsection*{The Classical Ground State}

In the presence of the harmonic trap, the classical equation of motion for $Z$ reads
\be i {\cal D}_t Z = \omega Z\label{zeom}\ee
There is a unique time independent solution, with $\dot{Z}=0$, obeying $[a_0,Z]=\omega Z$. 
%
%
This can also be viewed as the statement that rotating the phase of $Z$ is equivalent to a gauge transformation.
%
There is a unique solution to this equation and the constraint \eqn{constraint}  given by \cite{alexios},
%
%
%
\be Z_0 = \sqrt{\frac{k}{\pi\mu}}\left(\begin{array}{cccccc}  0\ & 1\ &  & &  \\  & 0\ & \sqrt{2} & &  \\   & & & \ddots &   \\ &  &  & 0 & \sqrt{n-1} &  \\  & & & & 0  \end{array}\right)\ \ \ {\rm and}\ \ \ \varphi_0 = \sqrt{k} \left(\begin{array}{c} 0 \\  0 \\ \vdots \\ 0\\ \sqrt{n}\end{array}\right)\label{sol}\ee
with $a_0 =  \omega\,{\rm diag} (n-1, n-2,\ldots, 2,1,0)$.

\para
As promised, $Z_0$ is not approximately diagonal. This reflects the fact that individual vortices do not have well-defined positions. Nonetheless, 
we can reconstruct a number of simple properties of the vortex solution from this matrix. The radius-squared of the disc can be thought of as the maximum eigenvalue of $Z_0^\dagger Z_0$ \cite{alexios}. To leading order in the vortex number $n$, this gives
\be R^2  \approx \frac{kn}{\pi\mu}\nn\ee
which agrees with our the radius of the classical vortex solution \eqn{R}. Meanwhile, the angular momentum of a given solution is  ${\cal J}={\rm Tr}\,Z^\dagger Z$. The  angular momentum of the ground state is 
\be {\cal J}_0 = \pi\mu\,{\rm Tr}\,\left(Z_0^\dagger Z_0\right) = \frac{kn(n-1)}{2}\label{j0}\ee
which, to leading order in $1/n$,  agrees with the angular momentum of the classical vortex solution \eqn{jmin}. 

\subsubsection*{The Quantum Ground State}

The quantisation of the matrix model \eqn{qhmm} was initiated in \cite{alexios} and explored in some detail in \cite{hellvram} and \cite{ks}. The individual components of the matrix $Z$ and vector $\varphi$ are promoted to quantum operators, with commutation relations
\be \pi\mu\,[Z_{ab},Z^\dagger_{cd} ] = \delta_{ad}\delta_{bc}\ \ \ {\rm and}\ \ \ [\varphi_a,\varphi_b^\dagger]= \delta_{ab}\nn\ee
We choose the vacuum state $|0\rangle$ such that $Z_{ab}|0\rangle = \varphi |0\rangle=0$. However  this does not, in general,  correspond to the ground state of the theory because the physical Hilbert space must obey the quantum version of the Gauss' law constraint \eqn{constraint}. It is useful to view the trace and traceless part of this constraint separately. The trace  constraint reads
\be \sum_{a=1}^n \varphi_a \varphi^\dagger_a = kn \ \ \ \Rightarrow\ \ \  \sum_{a=1}^n \varphi^\dagger_a \varphi_a = (k-1)n\label{traceconstraint}\ee
This means that physical states must have $(k-1)n$ $\varphi$-excitations. Note that the ordering of the original constraint has resulted in a shift $k\rightarrow k-1$. This will prove important below. 

\para
Meanwhile, the traceless part of the constraint \eqn{constraint} tells us that physical states must be $SU(n)\subset U(n)$ singlets. We can form such singlet operators out of $Z^\dagger$ and $\varphi^\dagger$ either from baryons or from traces. The baryonic operators are
\be \epsilon^{a_1\ldots a_n} (\varphi^\dagger Z^{\dagger\,p_1})_{a_1}\ldots(\varphi^\dagger Z^{\dagger\, p_n})_{a_n}\nn\ee
where $p_1,\ldots p_n$ are, necessarily distinct, integers. The trace operators are
\be {\rm Tr} (Z^{\dagger\,p})\nn\ee
There can be complicated relations between the baryonic and trace operators; explicit descriptions for low numbers of vortices were recently presented in \cite{amirak}.

\para
The trace constraint \eqn{traceconstraint} means that physical states contain exactly $k-1$ baryonic operators.  The harmonic trap endows these with an energy proportional to the number of $Z^\dagger$ excitations,
\be H= \omega{\cal J} = \omega\pi\mu\,\sum_{a,b=1}^nZ^\dagger_{ab}Z_{ba}\nn\ee
To minimise this energy, we must act with $k-1$ baryonic operators, each  with $p_i = i-1$. This results in the ground state
\be |{\rm ground}\rangle_k = \left[\epsilon^{a_1\ldots a_n} \varphi^\dagger_{a_1}(\varphi^\dagger Z^\dagger)_{a_2} 
\ldots (\varphi^\dagger Z^{\dagger\,n-1})_{a_n}\right]^{k-1} |0\rangle\label{ground}\ee
The angular momentum of this ground state coincides with that of the classical ground state \eqn{j0}.

\para 
There is a close resemblance between these ground states and the Laughlin states \cite{laughlin} for $n$ electrons at filling fraction $\nu=1/k$,
\be |{\rm Laughlin}\rangle_k= \prod_{a<b} (z_a-z_b)^k \, e^{-\frac{B^{\rm ext}}{4}\sum |z_a|^2} = \left[\epsilon^{a_1\ldots a_n} z_{a_1}^0 z_{a_2} \ldots z_{a_n}^{n-1}\right]^k e^{-\frac{B^{\rm ext}}{4}\sum |z_a|^2}\ \ \ \ \ \label{laughlin}\ee
A formal map between the states was suggested in \cite{hellvram}. A more rigorous study, which we now review, was provided in \cite{ks} (see also \cite{cappelli}). The first step is to identify appropriate coordinates on the phase space of the matrix model. There are a number of different choices, none of which have preferred status. Here we use the coherent state representation suggested in \cite{ks}. We diagonalise $Z$ and use its eigenvalues as coordinates on the phase space. (Non-diagonalisable matrices have zero measure.)

\para
The first result of \cite{ks} is that the $k=1$ ground state, $|{\rm ground}\rangle_1 = |0\rangle$, is precisely the $\nu=1$ Laughlin state describing a filled Landau level,
\be |0\rangle = |{\rm Laughlin}\rangle_1\nn\ee
For $k>1$, the map to the Laughlin wavefunction is not exact. Instead, the wavefunctions agree only at large separation
\be |{\rm ground}\rangle_k \rightarrow |{\rm Laughlin}\rangle_k\ \ \ {\rm for}\ \ \ |z_a-z_b| \gg 1/\pi\mu\nn\ee
However, the matrix model states $|{\rm ground}\rangle_k$ differ from the Laughlin states as the particle approach: the wavefunctions still vanish as $z_a\rightarrow z_b$, but not with the same power.

\para
As we mentioned, there is nothing privileged about the choice of coordinates used above. One may try a different set of coordinates and see if there is better short-distance agreement with the Laughlin wavefunction. Indeed, other coordinates were suggested in \cite{ks,cappelli}, although none of them provide an exact match to  the Laughlin wavefunctions.

\para
The connection to vortices sheds some light on this. Because vortices are extended objects, there is no ``correct"  way to specify their positions as they approach. Correspondingly, it is not obvious that their physics is captured by a wavefunction describing point particles. Instead, the important questions are those which are independent of the choice of coordinates. The fact that the long-distance correlations in the matrix model ground states \eqn{ground} coincide with those of the Laughlin wavefunction suggests that these states describe the same universality class of quantum Hall fluids. In the rest of this paper, we show that this is indeed correct. We show that excitations of the matrix model describe chiral edge modes and quasi-holes. In particular, the latter have charge $1/k$ and fractional statistics, in agreement with the excitations of the Laughlin wavefunction.

\subsection{Edge Modes}\label{edgesec}

The classical excitations of the matrix model were described in \cite{alexios}. There are edge excitations of the droplet and there quasi-hole excitations although, for finite $n$,  there is no clear distinction between these. There are no quasi-particle excitations which, given the spacetime picture in terms of vortices, is to be expected. 
We first study the edge modes and show that they form a chiral boson.

\para
 The linear perturbations of the solution \eqn{sol}, consistent with the constraint \eqn{constraint}, were given in \cite{alexios}: they are remarkably simple,
\be \delta_l Z = (Z_0^\dagger)^{l-1} \ \ \ {\rm and}\ \ \ \ \delta_l\varphi = 0 \ \ \  \ {\rm with}\ \ \ \ l=1,\ldots,n\label{pert}\ee
These were interpreted in \cite{alexios} as area-preserving deformations of the disc, restricted to the first $n$ Fourier modes. 

\para
We now show that the dynamics is that of a chiral, relativistic boson.  To do this, we write
\be Z(t)  = Z_0 + \sum_{l=1}^{n} c_l(t) Z_0^{\dagger\,l-1}\nn\ee
with complex coefficients $c_l$. Plugging this ansatz into the action \eqn{qhmm}, we have the following expression for the effective dynamics of $c_l$,
\be S =  \pi\mu\sum_{l,p=1}^n\int dt \ i  {\rm Tr}(Z_0^{l-1}Z_0^{\dagger\,p-1})\,c_l^\star \dot{c}_p+  \left[ {\rm Tr}(a_0\,[Z_0^{\dagger\,p-1},Z_0^{l-1}]) -  \omega  {\rm Tr}(Z_0^{l-1}Z_0^{\dagger\,p-1})\right]c_l^\star c_p \nn\ee
where we have dropped the constant contribution \eqn{j0}. We need to compute two traces, both involving $Z_0$ given in \eqn{sol}. The first is
\be \pi\mu\,{\rm Tr}\,Z_0^{l-1}Z_0^{\dagger\,p-1} \equiv \Theta_l\delta_{lp}\ \ \ {\rm with}\ \ \ \Theta_l= \frac{k^{l-1}}{l}n(n-1)\ldots(n-l+1)\nn\ee
The second trace involves $a_0$ and can be readily computed by invoking the relationship $\omega Z_0 = [a_0,Z_0]$,  to give $[a_0,Z_0^{\dagger\,p}]=-p\omega Z_0^{\dagger\,p}$. The action for the perturbations can then be written in the simple form,
\be S = \sum_{l=1}^n \Theta_l \int dt\ \left(i{c}_l^{\,\star}\,\dot{{c}}_l - \omega l {c}_l^{\,\star}{c}_l\right)\label{chiralact}\ee
This is the action for a real, chiral boson, defined on the edge of the Hall droplet. We parameterise the perimeter of the droplet by $\sigma \in [0,2\pi R)$ with $R$ given by \eqn{R}.  The continuum excitations then take the form
\be c(\sigma,t) = \frac{1}{\sqrt{2\pi}}\sum_{l=-\infty}^\infty {e^{il\sigma/R}}\sqrt{\frac{\Theta_l}{l}} \ c_l(t)\ \ \ \ \ \ {\rm with}\ \ \ c_{-l}=c_l^\star\nn\ee
Then the action \eqn{chiralact} becomes
\be S =  - \int dt d\sigma\ \partial_tc\,\partial_\sigma c + (\omega R) \partial_\sigma c \,\partial_\sigma c\nn\ee
This is the form of the action for a chiral boson proposed in \cite{jf}, now truncated to the lowest $n$ Fourier modes. 
The action  describes modes propagating in one direction around the disc with velocity  $v=\omega R$. A previous derivation of the chiral boson edge theory from the matrix model was given in \cite{edge}, albeit in a model with a different potential. It is unclear to us how that derivation relates to the one above. 

\para
Note that as $n$ increases, the radius of the disc \eqn{R} scales as $\sqrt{n}$, while the number of Fourier modes increases linearly with $n$. The density of modes therefore scales as $1/\sqrt{n}$, suggesting the existence of a  continuum $d=1+1$ dimensional limit as $n\rightarrow \infty$. 

%
%

\subsubsection*{The Noncommutative Description Revisited}\label{noncomsec}

The original motivation for the quantum Hall matrix model was to provide a finite $n$ regularisation of Susskind's non-commutative approach to quantum Hall fluids \cite{susskind}. Taking the $n\rightarrow \infty$ limit of the matrix model, one can effectively drop the field $\varphi$ and the constraint \eqn{constraint} becomes
\be [X^1,X^2]=  i\frac{2\pi  \mu }{k} = i\frac{eB^{\rm ext}}{k}\nn\ee
We interpret this as a non-commutative plane. Expanding the action around the state \eqn{sol} gives rise to a Chern-Simons theory on this non-commutative plane, with fields multiplied using the Moyal product \cite{susskind}. The perspective  offered here shows that this non-commutative theory provides 
a hydrodynamic description of  the dynamics of $n\rightarrow \infty$ BPS vortices.

\para
There is no harmonic trap introduced in the non-commutative Chern-Simons description. Because it arises from the expansion around \eqn{sol}, all perturbative excitations of the theory are edge modes of an infinitely large disc, now consigned to asymptotia. However, these perturbation excitations are not the end of the story. 
There are many other non-perturbative bulk excitations. These correspond to separating vortices or, as we will see in the next section, creating a hole in the fluid of vortices.  The non-commutative Chern-Simons theory is capturing these modes.

\para
However, we have already seen a different description of these modes from the perspective of the $d=2+1$ dimensional spacetime picture: they are the gapless, lowest Landau level of an interacting boson that we saw in Section \ref{phasessec}.  It appears that the  Chern-Simons theory on the non-commutative plane is an alternative description of this lowest Landau level physics. 


 \subsection{Quasi-Holes}\label{holesec}
 
Let us now return to a finite droplet of vortices. While the infinitesimal perturbations of the droplet describe edge modes, one can also consider finite deformations. Of course,  if we make a large enough finite perturbation, then the droplet will eventually fragment into its component vortices.  However, there are deformations for which the droplet retains its integrity, but with a hole carved out in the middle. These are the quasi-holes of the quantum Hall effect.

\para
There is a simple classical solution describing a quasi-hole placed at the centre of the vortex \cite{alexios}. It arises by integrating the $n^{\rm th}$ Fourer mode, 
\be Z = \sqrt{\frac{k}{\pi\mu}}\left(\begin{array}{cccccc}  0\ & \sqrt{1+q}\ &  & &  \\  & 0\ & \sqrt{2+q} & &  \\   & & & \ddots &   \\ &  &  & 0 & \sqrt{n-1+q} &  \\ \sqrt{q} e^{in\omega t} 
& & & & 0  \end{array}\right)\label{classicalhole}\ee
%
%
%
%
%
This obeys the constraint \eqn{constraint} and equation of motion \eqn{zeom} with $a_0 =  \omega\,{\rm diag} (n-1, n-2,\ldots, 2,1,0)$ and 
 $\varphi=\varphi_0$. 
 
\para
 This solution should be thought of as a deficit of magnetic field in the middle of the Hall droplet \cite{alexios} (see also \cite{mark}). In other words, it is a quasi-hole.  Using the maximum and minimum eigenvalues of $Z^\dagger Z$ as a proxy for the inner radius $R_1$ and the outer radius $R_2$ of this annulus, we find
\be R^2_1 \approx \frac{kq}{\pi\mu}\ \ \ {\rm and}\ \ \  R^2_2 \approx \frac{k(n+q)}{\pi\mu}\nn\ee
which is consistent with the magnetic flux quantisation \eqn{fluxnumber} if $B$ remains constant for $R_1<r<R_2$. We can subject this interpretation to a further test. The angular momentum of the matrix model solution is given by
\be {\cal J}  = \pi\mu\,{\rm Tr}\, Z^\dagger Z = \frac{kn^2}{2} + knq\nn\ee
But we can also compute the angular momentum of an annular vortex by the same kind of calculation we used in \eqn{jmin}. We find
\be {\cal J}  \approx - \frac{\mu}{2}  \int_{R_1}^{R_2}  dr\ 2\pi r^3 B = \frac{kn^2}{2}+knq\nn \ee
confirming the solution \eqn{classicalhole} as a classical quasi-hole.

\para
There are, presumably, more complicated classical solutions, describing quasi-holes displaced from the origin, rotating with frequency $\omega$. Rather than searching for these classical solutions, we will instead describe their quantum counterparts. 

\subsubsection*{Quantum Quasi-Holes}

We claim that the quantum state describing $m$ quasi-holes, located at complex coordinates $\eta_i$, $i=1,\ldots,m$, is  
\be |\eta_1, \ldots, \eta_m\rangle_k \ \propto\ \prod_{i=1}^m\ \det(Z^\dagger - \eta^\dagger_i)\ |{\rm ground}\rangle_k \label{quasihole}\ee
where we have allowed for a normalisation constant.

\para
Let us first motivate this ansatz. Multiplying by $\det(Z^\dagger - \eta^\dagger)$ is equivalent to taking one of the baryonic operators in the ground state \eqn{ground} and replacing each occurrence of $\varphi^\dagger Z^{\dagger\,p}$ by $\varphi^\dagger Z^{\dagger\,p} (Z-\eta)^{\dagger}$. Under the coherent state map of \cite{ks}, where the eigenvalues of $Z$ are used as coordinates, this gives 
\be |\eta_1,\ldots,\eta_m\rangle_k\ \rightarrow\ \prod_a (z_a-\eta) |{\rm Laughlin}\rangle_k\nn\ee
which is indeed the Laughlin wavefunction for quasi-holes.

\para
As we vary the positions $\eta_i$, the resulting states $|\eta_1,\ldots,\eta_m\rangle$ are not linearly independent. This reflects the fact that these holes are made from a finite number of underlying particles. Nonetheless, for $|\eta_i|<R$, with $R=\sqrt{kn/\pi\mu}$ the size of the quantum Hall droplet \eqn{R}, we expect the state to approximately describe $m$ localised quasi-holes. This interpretation breaks down as the quasi-holes approach the edge of the droplet. Indeed, the states degenerate  and become approximately the same for any value of $|\eta_i|\gg R$. We'll see the consquences of this below. 
 
\para
In the presence of a harmonic trap, the states \eqn{quasihole} are not energy eigenstates unless $\eta_i = 0$. Nonetheless, it is simple to check that the time-dependent states, $|e^{i\omega t }\eta_1,\ldots,e^{i\omega t}\eta_m\rangle_k$, in which the quasi-holes orbit the origin, solve the time-dependent Schr\"odinger equation. In what follows, we will compute the braiding of the time independent states \eqn{quasihole}.

%
%
%

\para
In the quantum Hall effect, the quasi-holes famously have fractional charge and fractional statistics. We now show this directly for the states \eqn{quasihole}. 
We follow the classic calculation of \cite{frank} in computing the Berry phase accumulated as quasi-holes move in closed paths. However, there is a technical difference that is worth highlighting. In the usual Laughlin wavefunction, the overlap integrals are too complicated to perform directly. Instead, one resorts to the plasma analogy \cite{laughlin}. This requires an assumption that a classical 2d plasma exhibits a screening phase.

\para
A second route to computing the braiding of quasi-particles is provided by the link to conformal field theories \cite{mr}, where it is conjectured to be equivalent to the monodromy of conformal blocks. The primary focus has been on the richer subject of non-Abelian quantum Hall states. Different approaches include \cite{notmoore} and \cite{gn1,gn2}, the latter once again relying on a plasma analogy. See also \cite{seidel} for an alternative approach to braiding.

\para
We will now show that the matrix model construction of the quasi-hole states \eqn{quasihole} seems to avoid these issues and a direct attack on the problem bears fruit. We compute the Berry phase explicitly without need of a plasma analogy.
%

\subsubsection*{Fractional Charge}

We start by computing the charge of the quasi-hole under the external gauge field. To do this, we consider a single excitation located at $\eta=re^{i\theta}$.  We then adiabatically transport the quasi-hole in a circle by sending $\theta\rightarrow \theta + 2\pi$. If the quasi-hole has charge $q_{\rm QH}$ then we expect that the wavefunction will pick up the Aharonov-Bohm phase $\Theta$ proportional to  the magnetic flux $\Phi$ enclosed in the orbit,
\be \Theta(r)= \Phi q_{\rm QH}  = \pi r^2 B^{\rm ext} q_{\rm QH}  = \frac{2\pi^2 \mu r^2}{e} q_{\rm QH}\label{abphase}\ee
where we've used the value of $B^{\rm ext} = 2\pi \mu/e$ computed in \eqn{bext}, with  $e$  the charge of a single vortex. There is a more direct expression for $\Theta$, arising   as the Berry phase associated to the adiabatic change of the wavefunction, 
\be \Theta(r) = -i\int_0^{2\pi} d\theta\ _k\langle \eta | \,  \frac{\partial}{\partial \theta}  | \eta \rangle_k \label{berry1}\ee
Our task is to compute this phase. From this we extract $q_{\rm QH}$.

\para
To do this, it will help to introduce some new notation. We define the states  $|\Omega_l\rangle_k$, with $l=0,\ldots,n-1$ 
\be |\Omega_l\rangle_k 
= [\epsilon^{a_1,\ldots,a_n} \varphi^\dagger_{a_1}(\varphi^\dagger Z^\dagger)_{a_2}
\ldots 
&&\!\!(\varphi^\dagger Z^{\dagger\,l-1})_{a_l}
(\varphi^\dagger Z^{\dagger\,l+1})_{a_{l+1}}
\ldots (\varphi^\dagger Z^{\dagger\,n})_{a_n} \nn\\
&&\  \ \ \ \ \ \  \left[\epsilon^{b_1,\ldots,a_n} \varphi^\dagger_{b_1}(\varphi^\dagger Z^\dagger)_{b_2} 
\ldots (\varphi^\dagger Z^{\dagger\,n-1})_{b_n}\right]^{k-2} |0\rangle \nn \ee
Each of these is an eigenstate of angular momentum, with ${\cal J} = {\cal J}_0 + \pi\mu(n-l)$.  We can expand the quasi-hole state \eqn{quasihole} in this basis as
\be |\eta\rangle_k \propto \sum_{l=0}^{n-1} (-\eta^\dagger)^l| \Omega_l\rangle_k \nn \ee
\begin{figure}[!h]
\begin{center}
\includegraphics[height=1.9in]{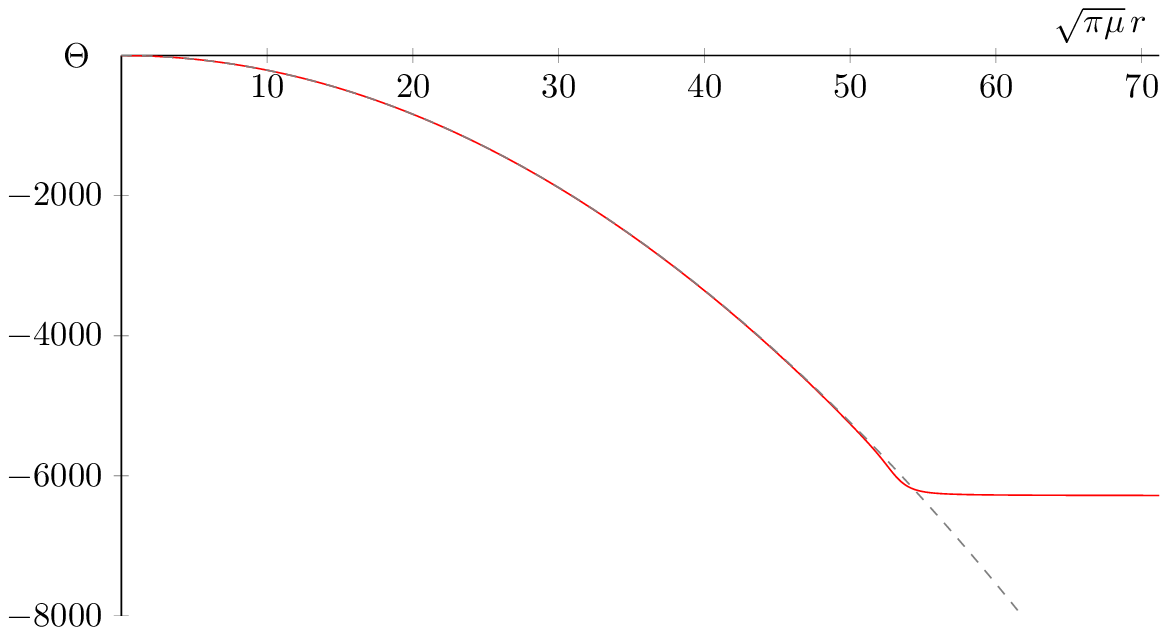}
\end{center}
\caption{The Berry phase for a single quasi-hole in $n=1000$ vortices with $k=3$. The phase $\Theta$ (solid, red)  and the expected phase for a particle of charge $-e/k$ in the field $B^{\rm ext}$ (gray, dashed) are both plotted.} \label{phaseflux}
\end{figure}
\noindent
Because the   $|\Omega_l\rangle_k$ have different angular momenta, they are orthogonal. We write their inner product as 
\be _k\langle\Omega_p|\Omega_l\rangle_k = \lambda(l;k)\,\delta_{lp}\nn\ee
In terms of these inner products, the Berry phase \eqn{berry1} is simply written as
\be \Theta(r) = 2\pi i\ \frac{\sum_{l=0}^n il \lambda(l;k)\, r^{2l}}{\sum_{l=0}^n \lambda(l;k)\, r^{2l}} \nn\ee
The computation of $\lambda(l;k)$ is not straightforward. (Indeed, this is the step in the usual calculation where one resorts to the plasma analogy.) We find the following result:
\be \lambda(l;k)  =  (\pi \mu)^{l-n} {n \choose l} \left[\prod_{a=0}^{n-l-1} (k a + 1) \right]  \ _k\langle {\rm ground} | {\rm ground}  \rangle_k \label{norm1}\ee
We relegate the proof of this statement  to Appendix \ref{norms}.

\para
Rather remarkably, the resulting sum can be written in closed form. We find
\be \Theta(r) = - 2 \pi^2 \mu r^2 \left(\frac{n}{(n-1)k + 1}\ \frac{\!_1 F_1(1-n, 2-n-1/k,\, {\pi \mu r^2}/{k})}{\!_1 F_1(\  -n, 1-n-{1}/{k},\, {\pi \mu r^2}/{k})}\right) \label{fphase} \ee
This is the ratio of confluent hypergeometric functions of the first kind.

\para
The result \eqn{fphase} is plotted in Figure \ref{phaseflux} for $n=1000$ vortices and $k=3$.  The plot shows clearly that, for $r<R$, the Berry phase $\Theta$ coincides with the expected Aharonov-Bohm phase \eqn{abphase} if the charge of the quasi-hole is taken to be
\be q_{\rm QH} = -\frac{e}{k}\nn\ee
This, of course, is the expected result \cite{laughlin, frank}.

\para
Our Berry phase computation also reveals finite size effects.  The magnitude of the Berry phase reaches a maximum of $2\pi n$ at  $r=R$, the edge of the droplet. Outside this disc, the Berry phase no longer increases and the picture in terms of  quasi-holes breaks down. One can also use the result above to determine the size of the edge effects; numerical plots reveal them to be small as long as $k\ll n$. 

\para
There is another interpretation of the quasi-hole state \eqn{quasihole}: it is an excitation of the fundamental boson $\phi$ in the  Hall phase \eqn{hallphase}. Now the Aharonov-Bohm phase arises because this particle has charge 1 under the statistical gauge field with magnetic field $B=-2\pi\mu/k$. This is a pleasing, dual perspective. The vortices are solitons constructed from $\phi$. But, equally, we see that we can  reconstruct  $\phi$ as a collective excitation of many vortices!

\subsubsection*{Fractional Statistics}

Let us next consider the statistics of quasi-holes as they are braided. To do this, we consider a state with two excitations, $|\eta_1,\eta_2\rangle_k$. It is simplest to place the first at the origin, $\eta_1=0$, and transport the second in a full circle. This is equivalent to exchanging the quasi-holes twice and computes double the statistical phase. Of course, there is also a contribution from the Aharonov-Bohm phase $\Theta(r)$ described above and we must subtract this off. The resulting statistical phase, $\Theta^{\rm stat}$, is then given by
\be 2\Theta^{\rm stat}(r) = -i\int_0^{2\pi} d\theta\ _k\langle 0, \eta | \,  \frac{\partial}{\partial \theta}   | 0, \eta \rangle _k\ - \Theta(r)\nn\ee
where, again $\eta = re^{i\theta}$. 

\para
To compute the statistical phase, we need yet more inner products. We define the states
\be | \Omega_{0,l} \rangle_k = [\epsilon^{a_1,\ldots,a_n}
(\varphi^\dagger Z^\dagger)_{a_1}
(\varphi^\dagger Z^{\dagger\,2})_{a_2}
\ldots\!\! && 
(\varphi^\dagger Z^{\dagger\,l})_{a_l}
(\varphi^\dagger Z^{\dagger\,l+2})_{a_{l+1}}
\ldots (\varphi^\dagger Z^{\dagger\,n+1})_{a_n} \nn\\
&&\ \ \ \ \  \ \ \left[\epsilon^{b_1,\ldots,a_n} \varphi^\dagger_{b_1}(\varphi^\dagger Z^\dagger)_{b_2} 
\ldots (\varphi^\dagger Z^{\dagger\,n-1})_{b_n}\right]^{k-2} |0\rangle \nn \ee

\begin{figure}[!h]
\begin{center}
\includegraphics[height=2.2in]{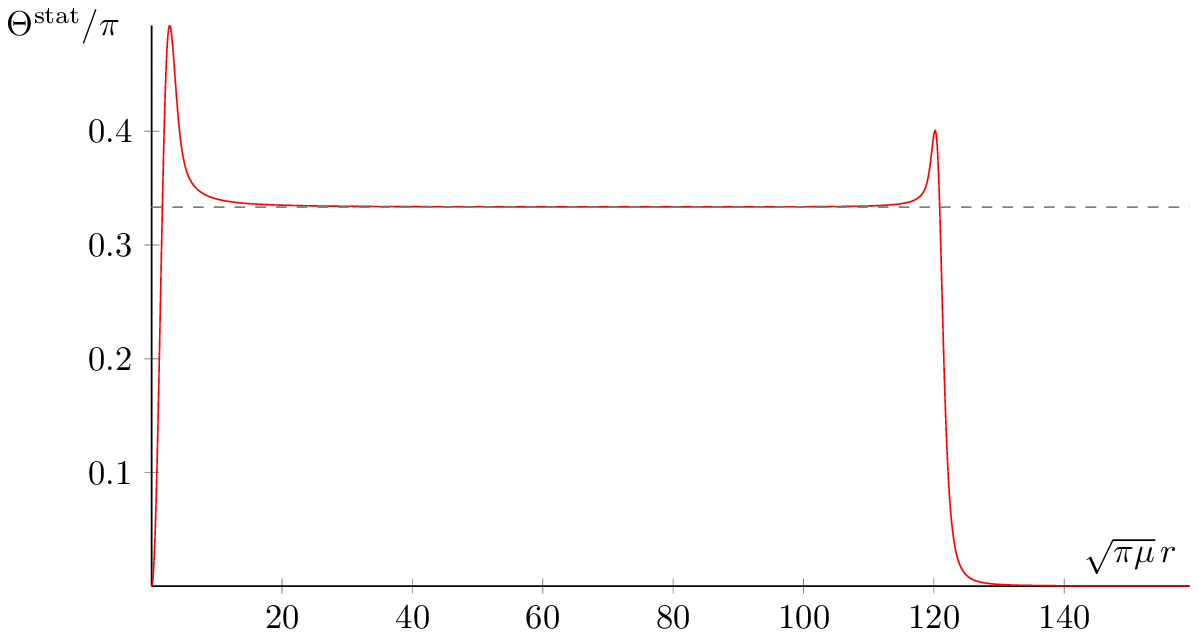}
\end{center}
\caption{The statistical phase for a quasi-hole encircling a second quasi-hole at the origin for $n=5000$ and $k=3$. The Berry phase $\Theta^{\rm stat}$ (solid, red) is plotted, together with the expected phase for a particle of statistics $\pi/k$ (gray, dashed).} \label{statflux}
\end{figure}
\noindent

This is similar to $|\Omega_l\rangle_k$, defined previously, except now each factor of $Z^\dagger$ has been increased by 1. This is the effect of placing the extra quasi-hole at the origin. (For more general locations of the quasi-hole, we would need the obvious generalisations of these states $|\Omega_{l',l}\rangle_k$.) The states $|\Omega_{0,l}\rangle_k$ are again orthogonal. This time, we find the norm is given by
\be \frac{_k\langle \Omega_{0,l}|\Omega_{0,l}\rangle_k}{_k\langle {\rm ground} | {\rm ground}  \rangle_k } &=& (\pi \mu)^{l-2n}  {n \choose l} \left[\prod_{a=0}^{n-l-1} (k a + 1) \right] \nn\\  &&\times \left[\prod_{a=0}^{l-1} \ (k a + 1) \right]\left[\prod_{a=l}^{n-1} \ (k a + 2) \right]   \label{norm2}\ee
With these functions, it is straightforward to determine an expression for the statistical phase in terms of a sum over $n$ states. Once again, this sum has a closed form, this time given using regularised hypergeometric functions by
\be 2\Theta^{\rm stat}(r)  = \frac{2 \pi^2 \mu r^2}{k} 
\left(n\ \frac{\!_2 \tilde{F}_2(1+ {1}/{k},1-n; 1 + {2}/{k}, 2-n- {1}/{k};\, {\pi \mu r^2}/{k})}{\!_2 \tilde{F}_2({1}/{k}, -n; \,{2}/{k},\,1-n-{1}/{k};\, {\pi \mu r^2}/{k})} \right) - \Theta_r \nn \ee
We plot this for $n=5000$ and $k=3$ in Figure \ref{statflux}. All other plots with $k\ll n$ have similar features. We see that there is clearly an intermediate, parametrically large regime, in which the pair of particles are both far from the edge of the disc and far from each other, where their exchange statistics are given by
\be \Theta^{\rm stat } = \frac{\pi}{k}\nn\ee
This is the expected result for a quasi-hole at filling fraction $\nu=1/k$. 

\noindent

\section{Future Directions}\label{moresec}

Supersymmetry has long proven a powerful tool to understand physics at strong coupling in relativistic systems. It is clear that if this power could be transported to the non-relativistic realm, then supersymmetry may be employed to say something interesting about open problems in condensed matter physics. 

\para
In this spirit, there have been a number of recent papers in which mirror symmetry (which can be viewed as an exact particle-vortex duality in $d=2+1$  interacting systems) has been explored  in the presence of external sources. This has been used to study impurities \cite{shamit1,kennyme}, non-Fermi liquids \cite{shamit2} and the physics of the lowest Landau level  \cite{shamit3}. It would be interesting to follow the fate of mirror pairs (or Seiberg duals) under the non-relativistic limit.

\para
The Laughlin physics described in this paper is, of course, just the tip of the quantum Hall iceberg. A long-standing open problem has been how to generalise the quantum Hall matrix model of \cite{alexios} to more general filling fractions such as the Jain heirarchy.  (See \cite{dontthinkso} for an attempt.) The perspective offered in this paper suggests a route. It is known that the most general Abelian quantum Hall state can be captured by the K-matrix approach \cite{wenzee}, with an effective field theory given by several coupled Chern-Simons fields
\be L =  \frac{1}{4\pi} K_{IJ} \epsilon^{\mu\nu\rho}A_\mu^I \partial _\nu A_\rho^J + \frac{e}{2\pi}A_\mu^{\rm ext} t_I\epsilon^{\mu\nu\rho}\partial_\nu A_\rho^I +  \ldots\nn\ee
It is  a simple matter to generalise this to a non-relativistic supersymmetric theory. However, the dynamics of vortices in these theories have not been well studied.  A matrix model for the vortex dynamics in these theories would presumably furnish a description of the most general Abelian quantum Hall states. (A matrix model for vortices in a class of theories with product gauge groups was proposed in \cite{kennyme,mina}.)

\para
Another obvious generalisation is to look at vortices in non-Abelian $U(N)$ gauge theories. These were introduced in \cite{me,yung}. The vortices now have an internal degree of freedom and the moduli space is given by
\be \pi \mu [Z,Z^\dagger] + \sum_{i=1}^N\varphi_i\varphi_i^\dagger  = k{\bf 1}_n\nn\ee
modulo $U(n)$ gauge transformations. This model was previously discussed in the context of quantum Hall physics in \cite{polywilson,kimura}.
It would be interesting to better understand what these states are and how they are related to the underlying non-Abelian Chern-Simons theories.

\para
Finally, we have restricted our attention in this paper to the case with chemical  potential $\mu\neq 0$. With $\mu=0$, the theory develops a non-relativistic super-conformal symmetry. An index was proposed in \cite{nakindex} but, beyond this, little seems to be known about the structure of these theories. With non-relativistic, conformal theories playing an important role in various cold atom systems, these super-conformal theories appear worthy of further study.

\para
 We hope to return to some of these issues in the future.

\section*{Acknowledgements}

We would like to thank Nigel Cooper, Nick Dorey, Nima Doroud, Ami Hanany, Kimyeong Lee, Nick Manton and Andrew Singleton for useful comments and discussions. We are grateful to KIAS for their kind hospitality while this work was ongoing. We are supported by STFC and by the European Research Council under the European Union's Seventh Framework Programme (FP7/2007-2013), ERC grant agreement STG 279943, ``Strongly Coupled Systems".

\newpage
\appendix
\section{Appendix: Non-Relativistic Limits}\label{appa}

The Lagrangian \eqn{lag1} can be derived by starting from a relativistic Chern-Simons matter theory, with ${\cal N}=2$ supersymmetry, and taking a limit in which anti-particles decouple \cite{llm}.  A number of other non-relativistic theories with different gauge groups, and more supersymmetry, have been constructed in this manner \cite{nak1,nak2,nak3,kimyeong}. 

\para
In this appendix, we review this non-relativistic limit. We construct  a more general theory than that of \cite{llm}, with gauge group $U(N_c)$ and $N_f$ matter multiplets transforming in the fundamental representation of the gauge group. We also show how the chemical potential term $\mu$ can arise in this limit.

\para
We restrict our attention  to the bosonic fields and, only at the end, describe the generalisation to the fermions.  The bosonic Lagrangian for the ${\cal N}=2$ supersymmetric $U(N_c)$ Yang-Mills Chern-Simons theory is
\be S_{\rm rel} &=& -\int d^3x\ \frac{1}{4e^2}{\rm Tr}\,(F_{\mu\nu}F^{\mu\nu}) + \frac{k}{4\pi}\epsilon^{\mu\nu\rho}\, {\rm Tr}\,(A_\mu\partial_\nu A_\rho - \frac{2i}{3}A_\mu A_\nu A_\rho) + \frac{1}{2e^2}{\rm Tr}\,({\cal D}_\mu \sigma)^2 \nn\\ 
&&\ \ \ \ \ \ \ \ \ \ \ \ \ +\sum_{i=1}^{N_f} |{\cal D}_\mu \phi_i|^2 + \phi_i^\dagger \sigma^2 \phi_i + \frac{e^2}{2} {\rm Tr} \Big( \sum_i \phi_i \phi_i^\dagger - \frac{k\sigma}{2\pi} - v^2 \Big)^2\label{drain}\ee
Here $\sigma$ is the real, adjoint scalar which accompanies $A_\mu$ in the vector multiplet, while $\phi_i$ are fundamental scalars that live in chiral multiplets. We have included a Fayet-Iliopoulos term $v^2$, but not real masses for the $\phi_i$. This can be done and results in different inertial masses in the non-relativistic limit.

\para
Before proceeding, it's useful to perform some simple dimensional analysis. We work with $\hbar=1$. This, of course relates energy to inverse time scales. However, as we are ultimately interested in non-relativistic physics, we retain the speed of light $c$. This means that we have two dimensionful quantities, length $L$ and time $T$. 

\para
The factors of $c$ in \eqn{drain} are currently hidden in the notation. The measure is 
\be d^3x = c \,dt \,d^2x \label{measure}\ee
while the derivatives are
\be |{\cal D}_\mu \phi|^2 = -\frac{1}{c^2} |{\cal D}_t \phi|^2 + |{\cal D}_\alpha \phi|^2\nn\ee
with $\alpha =1,2$ indexing spatial directions. Similarly, $A_0 = A_t/c$. The action is dimensionless. The other fields have dimensions $[A_t] = T^{-1}$ and $[A_\alpha] = [\sigma] =  L^{-1}$ and $[\phi]=L^{-1/2}$. The parameters have dimension $[k]=0$ and $[e^2] = [v^2] = L^{-1}$.

\para
We first take the infra-red limit, $e^2\rightarrow \infty$, to remove the Yang-Mills term. This also imposes the D-term as a constraint:  
\be 
\frac{k\sigma}{2\pi} = \sum_i \phi_i\phi_i^\dagger - v^2\label{thisisit}\ee
Using this to integrate out the adjoint scalar $\sigma$,  the scalar  potential terms in \eqn{drain} become
\be V =  \left(\frac{2\pi}{k}\right)^2\sum_i {\rm Tr}\,\phi_i\phi_i^\dagger\Big(\sum \phi_j\phi_j^\dagger - v^2\Big)^2\label{6pot}\ee
This kind of sextic potential is standard in supersymmetric Chern-Simons theories. 
The next step is to take the non-relativistic limit by discarding anti-particle excitations. To this end, we make the ansatz
\be  \phi_i(x,t) = \frac{1}{\sqrt{2mc}} \tilde{\phi}_i(x,t) e^{-imc^2t}\label{nonrelphi}\ee
Here $m$ is the mass of $\phi$, which we read off from the quadratic term in the potential \eqn{6pot}
\be m  = \frac{2\pi v^2}{kc}\nn\ee
The key point of the non-relativistic limit is that $\tilde{\phi}$ varies much more slowly that the frequencies $mc^2$ set by the mass gap. In particular, this means that the ansatz \eqn{nonrelphi} prohibits anti-particle excitations which scale as $e^{+imc^2 t}$. Plugging the ansatz \eqn{nonrelphi} into the kinetic terms gives, after an integration by parts,
\be \frac{1}{c^2} |{\cal D}_t\phi|^2 = \frac{1}{2mc}\left(\frac{1}{c^2}|{\cal D}_t\tilde{\phi}|^2  + 2im \tilde{\phi}^\dagger {\cal D}_t\tilde{\phi} + m^2 c^2 |\tilde{\phi}|^2\right)\nn\ee
 The overall factor of $1/c$ is cancelled by the factor of $c$ in the measure \eqn{measure}.
The third term, $m^2 c^2 |\tilde{\phi}|^2$ is designed to cancel the quadratic term in the potential. We now take the non-relativistic limit $c\rightarrow \infty$. In doing so, we're left only with the term linear in time derivatives. We can repeat this for all other terms in the action. In particular, taking a similar scaling of the potential \eqn{6pot}, leaves us only with the quartic coupling
\be V = - \frac{\pi}{kmc} \sum_{ij} (\tilde{\phi}_j\tilde{\phi}_i)(\tilde{\phi}_i^\dagger\tilde{\phi}_j)\nn\ee
The same scaling can be applied to the fermions in the original ${\cal N}=2$ theory. The end result is a $U(N_c)$ Chern-Simons theory, coupled to $N_f$ fundamental matter multiplets. To describe it, we revert to  the notation $\tilde{\phi}\rightarrow \phi$. The final non-relativistic action is
\be S = \int dt d^2x \ &&  \sum_{i=1}^{N_f} i\phi_i^\dagger {\cal D}_t \phi_i + i \psi_i^\dagger {\cal D}_t \psi_i - \frac{k}{4\pi} {\rm Tr}\,\epsilon^{\mu\nu\rho}(A_\mu \partial_\nu A_\rho - \frac{2i}{3} A_\mu A_\nu A_\rho) \nn \\ && - \frac{1}{2m}\sum_{i=1}^{N_f} \left( {\cal D}_\alpha \phi_i^\dagger{\cal D}_\alpha \phi_i + {\cal D}_\alpha \psi_i^\dagger {\cal D}_\alpha \psi_i + \psi_i^\dagger B \psi_i\right) \nn\\ && - \frac{\pi}{mk}\sum_{i,j} \left[(\phi_j^\dagger\phi_i)(\phi_i^\dagger\phi_j) - (\phi_j^\dagger\psi_i)(\psi_i^\dagger\phi_j) + 2 (\phi_i^\dagger\phi_j)(\psi_j^\dagger \psi_i)\right]\label{nonrelact}\ee
For $U(1)$ with $N_f=1$, this is the action \eqn{lag1} when $\mu=0$. (We have used the notation $\partial_0$ rather than $\partial_t$ in \eqn{lag1}.)

\para
The action \eqn{nonrelact} is invariant under superconformal transformations described in \cite{llm} and, in more detail, in \cite{nakindex}. The generators of the supersymmetries are the obvious non-Abelian generalisations of \eqn{Q1} and \eqn{Q2}.

\subsubsection*{Adding a Chemical Potential}

The action \eqn{lag1} also includes a chemical potential $\mu$ which plays a crucial role in our quantum Hall story. It is straightforward to add an analogous to term to the  relativistic Lagrangian \eqn{drain}. It is
\be {\cal L}_\mu  = \mu {\rm Tr}\,(A_0 - \sigma)\label{addmu}\ee
Obviously this breaks $d=2+1$ Lorentz invariance. It preserves two of the four supercharges. Indeed, such terms are well known in the context of quantum mechanics models with ${\cal N}=(0,2)$ supersymmetry and were first introduced in \cite{evanati}. In taking the infra-red limit, the $\sigma$ term in \eqn{addmu} gets replaced by $\sum \phi_i\phi_i^\dagger$ through the constraint \eqn{thisisit}. The resulting interaction terms of the non-relativistic theory are
\be
 V =  \frac{\pi}{mk}\sum_{i,j} \left[(\phi_j^\dagger\phi_i)(\phi_i^\dagger\phi_j)  - \mu \phi_i^\dagger \phi_i - (\phi_j^\dagger\psi_i)(\psi_i^\dagger\phi_j) + 2 (\phi_i^\dagger\phi_j)(\psi_j^\dagger \psi_i)\right]\nn\ee
Despite the fact that the relativistic theory with the deformation \eqn{addmu} preserves only one complex supercharge, both supercharges \eqn{Q1} and \eqn{Q2} are recovered after taking the non-relativistic limit. However, as we have seen, only $Q_2$ remains a symmetry of the spectrum.


\section{Appendix: The Geometry of the Vortex Moduli Space}\label{appgeom}

In this appendix, we review a few basic facts about the geometry of the vortex moduli space. Suppose that we have at our disposal the most general solution to the vortex equation with winding $n$,
\be \phi(x;X)\ \ \ {\rm and}\ \ \ A_z(x;X)\nn\ee
We define $2n$ {\it zero modes} $(\delta_a\phi,\delta_a A_z)$ to be the infinitesimal deformations which take us from one solution to another. 
\be \delta_a \phi = \frac{\partial \phi}{\partial X^a} +i \alpha_a \phi\ \ \ {\rm and}\ \ \ \delta_aA_z=\frac{\partial A_z}{\partial X^a} +\partial_z\alpha_a\label{zeromodes}\ee
Here $\alpha_a(x;X)$ is an accompanying gauge transformation. By construction, these zero modes solve the linearised versions of the vortex equations \eqn{vortex} for any choice of $\alpha(x,X)$. This ambiguity is fixed by further requiring that the zero modes obey the  background gauge condition,
\be \partial_z\,\delta_a  A_z + \partial_{\bar{z}}\,\delta_a A_{\bar {z}} =  \frac{2\pi}{k}\left(i \phi\delta_a\phi^\dagger-i\phi^\dagger \delta_a \phi\right) \label{bgauge}\ee
The metric on the vortex moduli space ${\cal M}_n$ is constructed by taking the overlap of the zero modes
\be g_{ab} = \int d^2x\ \frac{k}{\pi}\left(\delta_a A_{\bar{z}}\,\delta_b A_z + \delta_a A_z\,\delta_b A_{\bar{z}}\right) + \left(\delta_a\phi^\dagger \,\delta_b\phi + \delta_a\phi\,\delta_b\phi^\dagger\right)\label{metric}\ee
In relativistic theories, this metric plays an important role: the low-energy dynamics of the vortices is described by a sigma-model on ${\cal M}_n$ with metric $g_{ab}$. The metric is known to be free of singularities. It is also K\"ahler, inheriting its complex structure  from the natural action of complex conjugation on the fields. The associated K\"ahler form is
\be \Omega_{ab} = i\int d^2x\ \frac{k}{\pi}\left(\delta_a A_{\bar{z}}\,\delta_b A_z - \delta_a A_z\,\delta_b A_{\bar{z}}\right) + \left(\delta_a\phi^\dagger\, \delta_b\phi - \delta_a\phi\,\delta_b\phi^\dagger\right)\label{kahlerform}\ee
We now show that this K\"ahler form governs the first order dynamics of vortices in our model.  We will prove that the effective action for vortices is given by
\be S_{\rm vortex}   = \int dt\ {\cal F}_a(X)\dot{X}^a\ \ \ \ {\rm with} \ \ \ \ d{\cal F} = \Omega\nn\ee
This result was previously derived in \cite{manton,romao,nunomartin}.

\para
We work in the usual spirit of the moduli space: we promote the collective coordinates of the static solutions to be time dependent: $X^a(t)$. We then substitute this time-dependent ansatz into the kinetic terms of the action \eqn{lag1}. This results in an effective vortex action,
\be S = \int d^3x  \frac{ik}{2\pi}\left(A_{\bar{z}}\dot{A}_z-A_z\dot{A}_{\bar{z}} \right)+ \frac{i}{2}\left( \phi^\dagger\dot{\phi} - \dot{\phi}^\dagger\phi\right)  \equiv  \int dt\, {\cal F} _a(X) \dot{X}^a\label{kinaction}\ee
with
\be {\cal F} _a(X) = \frac{i}{2}\int d^2x\ \frac{k}{\pi}\left(A_{\bar{z}}\frac{\partial A_z}{\partial X^a} - \frac{\partial A_{\bar{z}}}{\partial X^a}A_z\right) 
+ \left(\phi^\dagger\frac{\partial \phi}{\partial X^a} - \frac{\partial \phi^\dagger}{\partial X^a}\phi\right)\nn\ee
Note that the kinetic terms in \eqn{kinaction} contain time derivatives rather than covariant time derivatives. This is because the $A_0$ terms in \eqn{lag1} multiply Gauss' law and so necessarily vanish. Correspondingly, the expression for ${\cal F} _a$ above contains partial derivatives of fields which differ from the zero modes defined in \eqn{zeromodes} as they are missing the contribution from the gauge transformation $\alpha_a(x;X)$.

\para
The 2-form $\tilde{\Omega} = d{\cal F} $ is
%
%
%
\be \tilde{\Omega}_{ab} = \frac{\partial{\cal F} _a}{\partial X^b} - \frac{\partial {\cal F} _b}{\partial X^a}
 = i\int d^2x\ \frac{k}{\pi}\left(\frac{\partial A_{\bar{z}}}{\partial X^a}\,\frac{\partial A_z}{\partial X^b} - \frac{\partial A_z}{\partial X^a}\,\frac{\partial A_{\bar{z}}}{\partial X^b}\right) + \left(\frac{\partial \phi^\dagger}{\partial X^a}\, \frac{\partial \phi}{\partial X^b} - \frac{\partial \phi}{\partial X^a}\,\frac{\partial \phi^\dagger}{\partial X^b}\right)\nn\ee
Our goal is to show that $\tilde{\Omega}_{ab} = \Omega_{ab}$, the K\"ahler form defined in \eqn{kahlerform}. The expressions look similar. They differ because the expression for $\Omega_{ab}$ includes  extra contributions from the gauge fixing terms. We now show that these terms vanish. 

\para
The proof is very similar to that given recently in \cite{andrew} in the context of first order motion on the instanton moduli space.  We take the difference
\be \Omega_{ab} - \tilde{\Omega}_{ab} &=& i\int d^2x\ \frac{k}{\pi} \left(\frac{\partial A_{\bar{z}}}{\partial X^a} \partial_z\alpha_b 
 -\frac{\partial A_z}{\partial X^a}\partial_{\bar{z}}\alpha_b \right) + \left(i\frac{\partial \phi^\dagger}{\partial A^a}\alpha_b \phi + i \phi^\dagger \alpha_a \frac{\partial \phi}{\partial X^a}\right)  - \left( a\leftrightarrow b\right)\nn\\
\nn\\ &=&-\int d^2x \ \alpha_b \,\frac{\partial}{\partial X^a}\left(- \frac{k}{2\pi} B + \phi^\dagger \phi\right) - \left(a \leftrightarrow b\right) \nn\ee
%
where we have integrated by parts to get to the second line. But the term in brackets vanishes, courtesy of the Gauss' law constraint \eqn{vortex}. We learn that $d{\cal F} =\Omega$, the K\"ahler form, as advertised. 
Note that the proof above did not need us to use the background gauge fixing condition \eqn{bgauge}. While the metric \eqn{metric} is sensitive to the background gauge condition, the K\"ahler form \eqn{kahlerform} is not.


\section{Appendix: Overlap of Matrix Model States}\label{norms}

Our derivation of the fractional charge and statistics of quasi-holes relied on expressions for the norms of matrix model states given in \eqn{norm1} and \eqn{norm2}. These results have been derived previously, most notably in the context of the Calogero-Sutherland-Moser model. Because these results are stated in a slightly different language, we use this Appendix to explain the connection.

\para
The quantum Hall matrix model is well known to be equivalent to the bosonic integrable Calogero-Sutherland-Moser model \cite{alexios,alreview}. This describes identical particles in one spatial dimension, placed in a harmonic trap and  interacting via a specific inverse-square potential.  To see the connection we  begin, following \cite{ks2}, by working with a  coherent state representation of all matrix model states. Firstly, expand $Z = X_1 + i X_2$ into Hermitian and anti-Hermitian parts, and let the overcomplete states $\left|X,\phi\right>$ be defined by
\be \hat{X}_1 \left|X,\phi\right> = X \left|X,\phi\right>, \qquad \hat{\varphi} \left|X,\phi\right> = \phi \left|X,\phi\right> \nn \ee
together with the normalisation
\be \int e^{-\bar{\phi}\phi} d\phi d\bar\phi \prod_{i,j}dX_{ij} \left|X,\phi\right>\left<X,\phi\right| \equiv 1 \nn \ee
where we have added hats to emphasise which symbols denote the quantum operators. With respect to these states, we can write all states in terms of a wavefunction by taking inner products with $\left<X,\phi\right|$. This in turn gives us a way to compute the inner products of matrix model states by computing integrals over $X,\phi$. In what follows, we work with convention $\pi\mu=1$.

\para
On these wavefunctions, $Z^\dagger$ has the representation 
\be Z^{\dagger}_{ij} \equiv 2^{-1/2} \left(X_{ij} - \frac \partial {\partial X_{ji}} \right) \nn \ee
analogous to the raising operator of the more familiar Hermite polynomials. 
Hence, up to an overall normalisation, the states we are interested in all have wavefunctions of the form
\be \Phi_f(X,\phi) = f(Z^\dagger) \left[\epsilon^{a_1\ldots a_n} \bar{\phi}_{a_1}(\bar{\phi} X)_{a_2} \ldots (\bar{\phi} X^{n-1})_{a_n}\right]^{k-1} e^{-\frac 1 2 {\rm Tr} X^2 }  \nn \ee
where $f$ is some homogeneous, gauge-invariant polynomial. Specifically, we have the following correspondence:
\be
\begin{array}{ll}
\left| \mathrm{ground}\right>_k: & f(B) = 1 \\
\left| \Omega_l\right>_k: & f(B) = B^{a_1}_{\ [a_1} B^{a_2}_{\ a_2} \cdots B^{a_{n-l}}_{\ a_{n-l}]} \\
\left| \Omega_{0,l}\right>_k: & f(B) = \det B \cdot B^{a_1}_{\ [a_1} B^{a_2}_{\ a_2} \cdots B^{a_{n-l}}_{\ a_{n-l}]}
\end{array}
\nn \ee
At given $n,k$, we will denote the state with a given choice of $f$ simply by $\left| f \right>$.

\para
One can evaluate the action of $f$ on the state to obtain instead
\be \Phi_f(X,\phi) = \tilde{f}(X) \left[\epsilon^{a_1\ldots a_n} \bar{\phi}_{a_1}(\bar{\phi} X)_{a_2} \ldots (\bar{\phi} X^{n-1})_{a_n}\right]^{k-1} e^{-\frac 1 2 {\rm Tr} X^2 }  \nn \ee 
where at leading order $\tilde{f}(B) \sim 2^{(\deg f)/2} f(B)$.

\para
Now the relationship to the states of the Calogero model is seen by performing a change of variables: diagonalise $X$ via $X = U D U^{-1}$, where $D_{ij} = x_i \delta_{ij}$. Defining the Vandermonde determinant
\be \Delta = \epsilon^{a_1 \cdots a_n} x_{a_1}^0 \cdots x_{a_n}^{n-1} = \prod_{i<j} (x_i - x_j) \nn \ee
one sees that the wavefunction becomes
\be \Phi_f(X,\phi) &= \tilde{f}(D) \cdot \Delta^{k-1} \ e^{-\frac 1 2 x^2 } \cdot \prod_i (\bar{\phi} U)_i^{k-1} \nn \ee
Note that $\tilde{f}(D) \equiv \tilde{f}(x)$ is simply a polynomial in $x$, whose leading behaviour we can determine from $f$. Also, we can see that $U, \phi$ have decoupled from $x$.

\para
Hence, taking account of the Jacobian $\Delta^2$ for our change of variables, at a given $n,k$ all inner products satisfy
\be \left<f | g \right> = c_{n,k}  \int d^n x \ e^{-x^2} \Delta^{2k} \tilde{f}(x) \tilde{g}(x) \label{innerprod} \ee
where $c_{n,k}$ is a calculable constant which we do not need for our computation.

\para
As is shown in detail in \cite{ks2}, the key observation now is that the action of the matrix model Hamiltonian $H$ on our wavefunctions is given by $H \equiv \Delta^{-1} H_{\rm Cal} \Delta$, where $H_{\rm Cal}$ is the Hamiltonian of the Calogero model at statistical parameter $k$.
But the eigenstates of the Calogero model are known; they correspond precisely to the Hi-Jack polynomials, the multi-variable generalisations of the Hermite polynomials which are orthogonal with respect to the measure in \eqn{innerprod}. These are labelled by partitions $\lambda$.

\para
One may readily check that in fact $\tilde{f}, \tilde{g}$ in equation \eqn{innerprod} must be multiples of the generalised Hermite polynomials of Section 3 discussed in \cite{bakerforrester}. But now we can refer to Proposition 3.7 of that paper which is readily unpacked to give the ratios between the norms of general states. Concretely, their $H_\lambda(x)$ have leading term 
\be
H_\lambda(x) \sim 2^{|\lambda|} \frac { (x_1^{\lambda_1} x_2^{\lambda_2} \cdots x_n^{\lambda_n} + \mbox{distinct permutations})}{\mbox{number of distinct permutations}}
\nn \ee
and norms
\be \frac{\int H_\lambda^2(x) \ d\mu(x)}{\int d\mu(x)} = %
2^{|\lambda|} %
\prod_{(c,d) \in \lambda} %
\frac{ %
	(k \, l_\lambda(c,d)+(a_\lambda(c,d)+1))(k(l_\lambda(c,d)+1)+a_\lambda(c,d)) %
}{ %
	k(n-(c-1))+(d-1) %
} \nn \ee
Here, $|\lambda| = \sum_i \lambda_i$ is the number of cells in the corresponding Young diagram, and $a_\lambda(c,d), l_\lambda(c,d)$ are respectively the arm and leg length of the cell with coordinates $(c,d)$ in that diagram.

\para
All that remains is to work out what choice of $\lambda$ and normalisation correspond to the examples of $f$ given above for the matrix model states. It is easily found that
\be
\begin{array}{ll}
\left| \mathrm{ground}\right>_k: & \tilde{f} = H_{(0,0,\ldots,0)} \\
\left| \Omega_l\right>_k: & \tilde{f} = 2^{(n-l)/2} {n \choose l} H_{(1,1,\ldots,1,0,0,\ldots,0)} \\
\left| \Omega_{0,l}\right>_k: & \tilde{f} = 2^{(2n-l)/2} {n \choose l} H_{(2,2,\ldots,2,1,1,\ldots,1)}
\end{array}
\nn \ee
where there are $n-l$ instances of 1 (resp. 2) in the second (resp. third) partition and then \eqn{norm1} and \eqn{norm2} both follow on evaluating the above product.

\para
It is very clear how this generalises to arbitrary states in the matrix model, especially if one realises the close relationship between the partition $\lambda$ and the original definition of the matrix model states $\left| \Omega_l\right>_k$ and $\left| \Omega_{0,l}\right>_k$.

\end{document}